\documentclass[twocolumn,amsmath,amssymb,nofootinbib]{revtex4}
\usepackage[english]{babel}
\usepackage[dvips]{graphicx}
\usepackage{enumerate}
\begin{document}
\title{Dark Matter Candidates: A Ten-Point Test}
\author{Marco Taoso$^{1,2}$}
\author{Gianfranco Bertone$^{2}$}
\author{Antonio Masiero$^{1}$}
\affiliation{$^{1}$ INFN, Sezione di Padova, via Marzolo
8,Padova,35131,Italy} \affiliation{$^{2}$ Institut d'Astrophysique de Paris, UMR 7095-CNRS, Universit\'e Pierre et Marie Curie, 98 bis Boulevard Arago 75014, Paris, France}
\begin{abstract}
An extraordinarily rich zoo of non-baryonic Dark Matter candidates
has been proposed over the last three decades. Here we present a
10-point test that a new particle has to pass, in order to be
considered a viable DM candidate: I.) Does it match the appropriate
relic density? II.) Is it {\it cold}? III.) Is it neutral? IV.) Is
it consistent with BBN? V.) Does it leave stellar evolution
unchanged? VI.) Is it compatible with constraints on
self-interactions? VII.) Is it consistent with {\it direct} DM
searches? VIII.) Is it compatible with gamma-ray constraints? IX.)
Is it compatible with other astrophysical bounds? X.) Can it be
probed experimentally?
\end{abstract}

\maketitle
\vskip 1cm

\section*{INTRODUCTION}

To identify the nature of Dark Matter is one of the most important
open problems in modern cosmology. Although alternative explanations
have been proposed in terms of modified gravity, the discrepancies
observed in astrophysical systems ranging from galactic to
cosmological scales appear to be better understood in terms of a
dark, yet undiscovered, matter component, roughly 6 times more
abundant than ordinary baryons in the Universe (see
Refs.~\cite{bertone review, bergstromRev} for recent reviews).

The possible connection of this exciting problem with New Physics
beyond the Standard Model has prompted the proliferation of Dark
Matter candidates, that are currently being searched for in an
impressive array of accelerator, direct and indirect detection
experiments. As our understanding of particle physics and
astrophysics improves, we accumulate information that progressively
reduces the allowed regions in the DM particles parameter space.

Here, we present
a 10-point test that new particles have to pass in order to be considered
good DM candidates. We will work under the assumption that a single
DM species dominates the DM relic density, while contribution
from other species is subdominant; it is straightforward to generalize the
discussion to the case of multi-component DM. Furthermore, we will consider a standard
$\Lambda$CDM cosmological model, although we discuss the consequences
of more general models, allowing for instance a non-standard expansion
history at the epoch of DM freeze-out.

Each of the following ten points, that represent {\it necessary}
conditions for a particle to be considered a good DM candidate,  will
be discussed in a dedicated section, where we will review the literature
on the subject and present the most recent results. In each section
we will discuss how robust the constraints are, especially for those
that heavily rely on astrophysical quantities such as the local DM
density and velocity distribution, or the extrapolation of DM profiles
at the center of galactic halos, often affected by
large uncertainties.

A particle can be considered a good
DM candidate only if a positive answer can be give to {\it all} the following points:

\begin{enumerate}
{\it \item Does it match the appropriate relic density?
\item  Is it cold?
\item  Is it neutral?
\item  Is it consistent with BBN?
\item Does it leave stellar evolution unchanged?
\item Is it compatible with constraints on self-interactions?
\item Is it consistent with {\it direct} DM searches?
\item Is it compatible with gamma-ray constraints?
\item Is it compatible with other astrophysical bounds?
\item Can it be probed experimentally?}
\end{enumerate}

The distinction between {\it gamma-ray constraints} and {\it  other
astrophysical bounds}, in points 8) and 9), is rather artificial,
and it simply reflects the privileged role of photons in
astrophysics, since they propagate along straight lines (unlike
charged particles), and they can be detected with better sensitivity
than, say, neutrinos. The fact of considering gamma-ray photons is
then due to the fact that the decay or annihilation of some of the
most common candidates falls in this energy range.

We also note that, strictly speaking, the last point is not really a {\it necessary}
condition, as DM particles could well be beyond the reach
of current and upcoming technology. However, measurable evidence
is an essential step of the modern scientific method, and a candidate
that cannot be probed, at least indirectly, would never be
accepted as the solution to the DM puzzle.

\section{DOES IT MATCH THE APPROPRIATE RELIC DENSITY?}
\label{sec:chapter one}

The analysis of the Cosmic Microwave Background (CMB) anisotropies
is a powerful tool to test cosmological models, and to extract the
corresponding cosmological parameters. For instance, the angular
position of the peaks in the power spectrum of temperature
anisotropies is a sensitive probe of the curvature of the Universe
(see e.g. \cite{Scott Smoot, Dodelson} for a review and a more
extended discussion). The power spectrum of CMB anisotropies is
fitted within the Standard Cosmological Model with a number or free
parameters that depends onto
the priors.\\
The best fit of the three years WMAP data, with a 6 parameters flat
$\Lambda$CDM model and a power-law spectrum of primordial
fluctuations, gives \cite{Spergel}
$$ \Omega_{b}h^2=0.0223^{+0.0007}_{-0.0009}  \mbox{    } \Omega_{M}h^2=0.127^{+0.007}_{-0.013}   $$
for the abundance of baryons and matter, respectively. The
normalized abundance $\Omega_i$ is defined as
$\Omega_{i}=\rho_{i}/\rho_{c},$ where $\rho_c$ is the critical
density, and the scaled Hubble parameter $h$ is defined as $H_0
\equiv 100h \mbox{ km} \mbox{ s}^{-1} \mbox{ Mpc}^{-1}.$ A
joint-likelihood analysis on a larger data-sets including, besides
WMAP3, also small scale CMB experiments (BOOMERang, ACBAR, CBI and
VSA), Large-Scale Structures (SDSS, 2dFGRS) and SuperNova
(HST/GOODS, SNLS), further strengthens the constraints to
\cite{http://lambda.gsfc.nasa.gov}
$$ \Omega_{b}h^2=0.0220^{+0.0006}_{-0.0008}  \mbox{    } \Omega_{M}h^2=0.131^{+0.004}_{-0.010}.   $$
Note that the baryonic density is consistent with the determination
from big bang nucleosynthesis \cite{Sarkar? vedi PDG}
$$ 0.017<\Omega_b h^2 <0.024 \mbox{ (95 \% CL)}.$$

For a new particle to be considered a good DM candidate, a
production mechanism that reproduce the appropriate value of the
relic density must exist. Moreover, to guarantee its stability, its
lifetime must exceed the present age of the Universe. Taking in
account the estimates of the Hubble Space Telescope Key Project
\cite{Hubble constant} and in agreement with the result derived by
WMAP, $H_0 = 72 \pm 3 \mbox{ (statistical)}\pm 7 \mbox{
(systematic)} \mbox{ km} \mbox{ s}^{-1} \mbox{Mpc}^{-1}, $ we
require a lifetime $\tau \gtrsim 4.3 \times 10^{17} \mbox{ s}.$

In many proposed extensions of the Standard Model of particle physics,
the stability of the DM particle is ensured by
imposing a symmetry that forbids the decay of DM into Standard Model
particles. For example, R-parity in Supersymmetry
models (SUSY) \cite{Rparity 1,Rparity 2}, K-parity in Universal
Extra Dimensions Models (UED) \cite{K parity}, and T-parity in
Little Higgs Models \cite{t parity}, prevent the lightest {\it new}
particle in the respective theories from decay (see for
example Ref.\cite{Jungman Kamionkowki Griest} for a detailed
discussion on SUSY DM and Ref.\cite{Profumo Review} for a review on
UED DM).\\

\subsection*{Thermal relics}

Among the best DM candidates, there is a class of particles called
WIMPs (for weakly interacting massive particles), that are thermal
relics and naturally achieve the appropriate relic density.

The scenario goes as follows: the WIMP is in thermodynamic equilibrium with the plasma in the
early Universe, and it decouples when its interaction rate drops below the
expansion rate of the Universe. For a non-relativistic particle at decoupling,
the number density over the
entropy density remains frozen, i.e. the thermal relic {\it freezes-out}.
The evolution of the number density of a generic species $\chi$ in
the Universe, is described by the following Boltzmann equation:
$$   \dot{n}_{eq} + 3Hn = -\langle \sigma_{ann}v \rangle \left[ n^2 -n_{eq}^2 \right].$$
The second term in the l.h.s of the equation takes into account the
dilution of the number density due to the expansion of the Universe.
$\langle \sigma_{ann} v \rangle $  is the thermal average of the
annihilation cross section times velocity and it is parametrized
with a non-relativistic expansion in powers of $v^2$, as: $ \langle
\sigma_{ann} v \rangle = a + b \langle v^2\rangle
+\mathcal{O}(\langle v^4 \rangle) \simeq a + 6b/x,$ with $x\equiv
m/T.$

$n_{eq}$ is the equilibrium density of WIMPs in the plasma at
temperature T and for a non-relativistic specie is given by $n_{eq }
= g(\frac{mT}{2\pi})^{3/2} e^{-\frac{m_{\chi}}{T}} $, where $g$
denotes the number of degrees of freedom of $\chi$ and $m_{\chi}$ is
the WIMP mass.

The Boltzmann equation can be solved integrating it in two extreme
regions, long before and long after the WIMP freeze-out (e.g. WIMP
decoupling), and matching then the solutions. Skipping the calculation
details, that can be reviewed e.g. in \cite{Kolb Turner}, the relic
density today for a generic WIMP $\chi$  is \cite{bertone review}:
\begin{eqnarray}
\Omega_{\chi} h^2 & \thickapprox &\frac{1.07 \times 10^{9} \mbox{
GeV}^{-1} }{M_{Pl} }\frac{x_f}{\sqrt{g_{*f}}}
\frac{1}{(a+3b/x_f)} \nonumber \\
& \thickapprox & \frac{3 \times 10^{-27} \mbox{
cm}^3\mbox{s}^{-1}}{\langle \sigma_{ann} v\rangle}. \label{eqn:
Omega}
\end{eqnarray}
where $g_{*f}$ counts the relativistic degrees of freedom at the
decoupling, $M_{Pl}$ is the Planck mass and $x_f \equiv
m_{\chi}/T_f$ with $T_f$ the freeze-out temperature. The last line
is an order of magnitude estimate and it shows that the relic
abundance of a non relativistic decoupled specie strictly depends on
the annihilation cross section at freeze-out \cite{Jungman
Kamionkowki Griest}. Furthermore, it has to be noticed that the
annihilation cross section, for a particle of given mass, has a
maximum, imposed by the partial wave unitarity of the S matrix,
$\langle \sigma_{ann} v\rangle_{max} \sim 1/m_{\chi}^2 $
\cite{Unitarity bound, Hui Ubound}. Thus, with the use of Eq.
\ref{eqn: Omega}, the requirement $\Omega_{M} h^2 \lesssim 1$
implies the following constraint on the mass of the DM particle,
also called "unitarity bound" \cite{Unitarity bound}
$$m_{DM} \lesssim 340 \mbox{ TeV}.$$
Applying the more stringent constraint, obtained by WMAP, the upper
bound on $m_{DM}$ becomes:
$$ m_{DM} \lesssim 120 \mbox{ TeV}.$$

\begin{figure*}[tb]
\includegraphics[width=6.5cm]{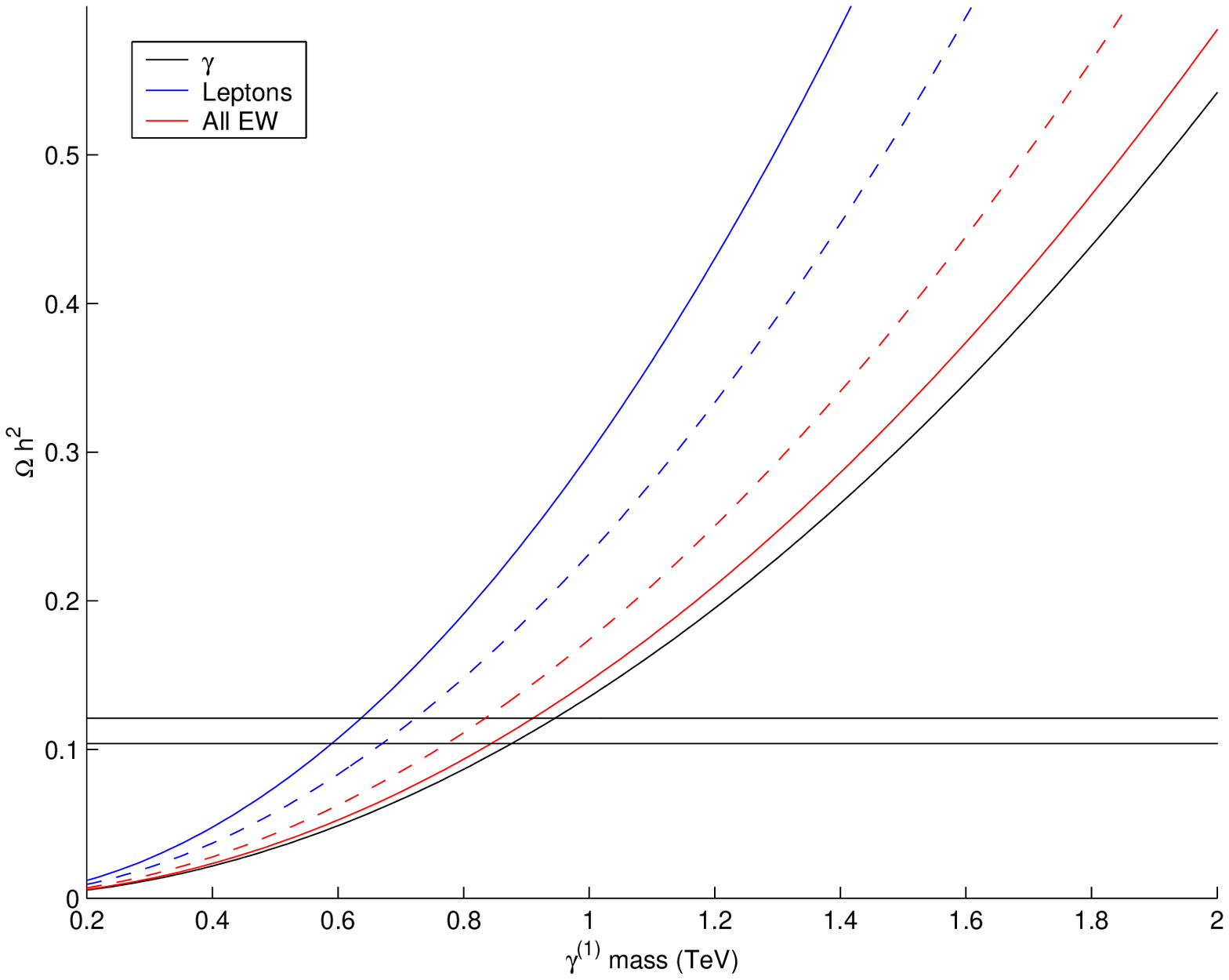}
\includegraphics[width=6.5cm]{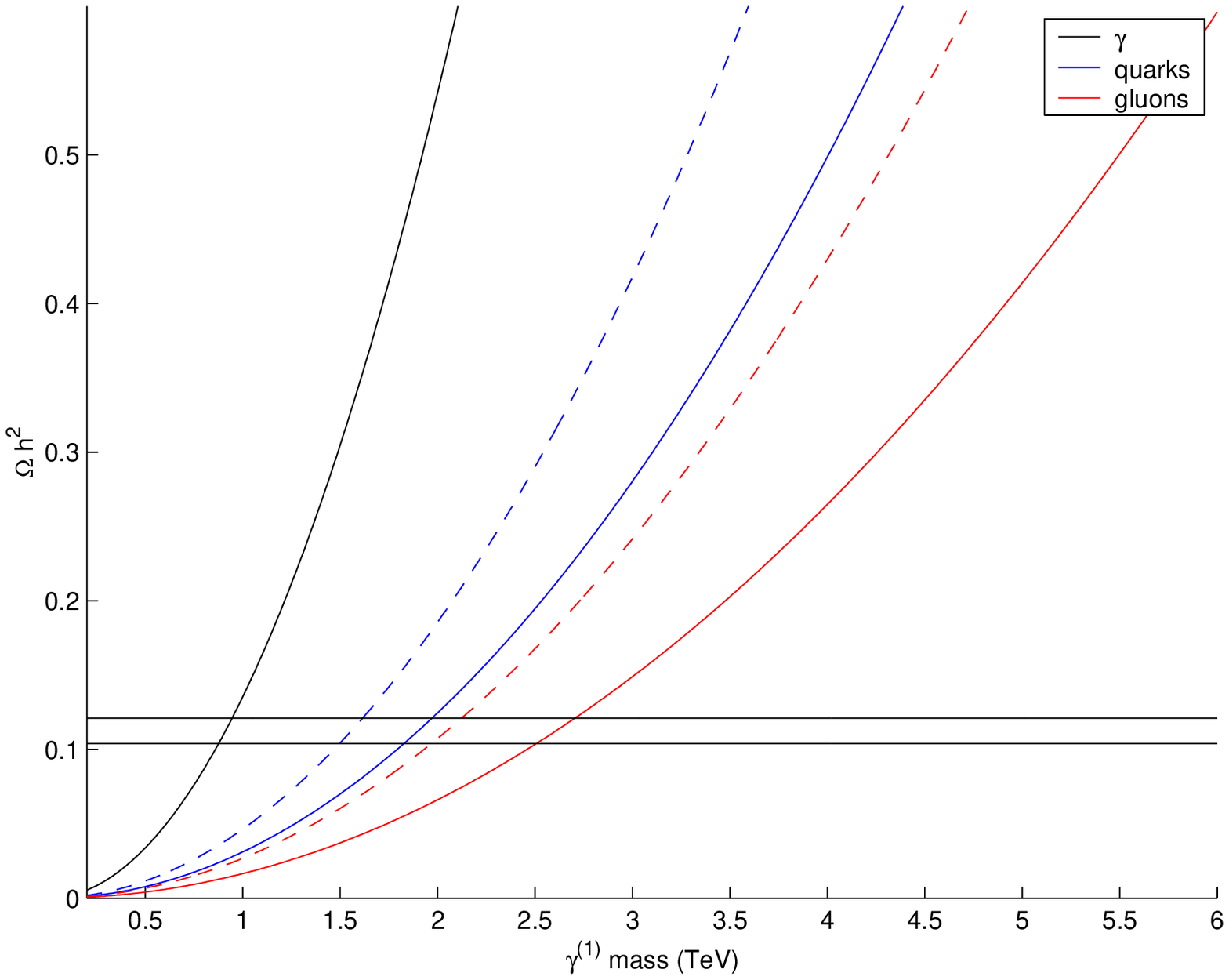}
\caption{Left: Relic Abundance of $B^{(1)}$ in the UED model as a
function of its mass after including no coannihilation (black line),
coannihilation with all leptons (blue) and all electroweak particles
(red). For the cases with coannihilation, the solid and dashed lines
are computed with a mass splitting $\delta = 0.01$ and $0.05$
respectively. Right: The same as in the left panel but accounting
for coannihilation of $B^{(1)}$ with all electroweak particles and
quarks (blue line), and all level-one KK particles, including KK
gluons (red line). Solid and dashed lines are for a mass splitting
$\delta = 0.01$ and $0.05$ respectively. From Ref.\cite{Burnell}.}
\label{fig:coannihilation}
\end{figure*}

However, this constraint was derived under the assumption that
particles were in thermal
equilibrium in the early universe, thus it
applies only to thermal relics, and can be evaded by species which are
non-thermally produced.

The standard computation of the thermal relic abundance discussed
above presents three important exceptions, as it has been shown,
following previous ideas \cite{prima griest}, by Griest and Seckel
\cite{Griest Seckel}. They take place for WIMPs lying near a mass
threshold, for annihilations near to a pole in the cross section, or
in presence of coannihilations. The last effect occur when a
particle that shares a quantum number with the WIMP, is nearly
degenerate in mass with it. If the mass gap is low enough (roughly
$\lesssim 10 \%$) the coannihilation reactions, involving WIMP
particles, can control the WIMP abundance and lower
or enhance it.\\
Full relic density calculations, including all coannihilations, have
been performed e.g. for the supersymmetric neutralino, for which
numerical codes such as DarkSUSY \cite{darksusy} and micrOMEGAs
\cite{micromega} are publicly available. Coannihilations have a
dramatic effect on the relic density, and they can lower it by a
factor of up to several hundreds (see e.g. \cite{Edsio}).


Coannihilations are also important in UED models, where the relic
density of the first excited state of the B, which may be the lightest
Kaluza-Klein particle (LKP) and a viable DM candidate, may be
enhanced or lowered depending on the coannihilation channel
\cite{Servant Tait, Burnell, Matchev}. See Fig.\ref{fig:coannihilation}.\\

Deviations from Standard Cosmology can substantially change the
picture. For example, due to the presence of scalar fields, the
universe may undergo a period of much higher expansion rate and the
relic abundance of a WIMP may result increased by several orders of
magnitude \cite{Salati, Masiero}. Furthermore, the production of
entropy in the Universe after the WIMP decoupling may dilute its
abundance, e.g. due to out-of-equilibrium decays of non-relativistic
particles or to first-order phase transitions
\cite{Kawasaki Entropy,Abazajan Koushiappas,Asaka Kusenko Shapo,Ellis Nanopoulos Entropy}.\\

It is also possible that DM particles did not experience the thermal
history depicted above, and that they have inherited the appropriate
relic density through the decay of a more massive species, that has
earlier decoupled from the thermal bath. This is e.g. the case for
SuperWeakly Interacting Massive Particles (SWIMP), such as the LSP
gravitino in SUSY and the first excitation of the graviton in UED,
which are produced by late decays of the next-to lightest particles
(NLSP/NLKP) in the respective theories \cite{Feng Rajaraman03,
EllisOliveSantoso04, Feng RefPRD03, FengRajaTakaPRD03} and whose
relic abundances are simply the rescaled thermal relic densities of
the NLPs:
$$\Omega_{\rm{SWIMP}} = \frac{m_{\rm{SWIMP}}}{m_{\rm{NLP}}} \Omega_{\rm{NLP}}.$$

Other production mechanisms may actually be concomitant for such
candidates, such as the production at reheating after the end of the
inflationary era (see e.g. \cite{Feng RefPRD03, Boltz
Brandeburg Bucch}. See also below for a brief discussion of gravitino production).\\

\subsection*{Other production mechanisms}

Very heavy DM candidates, such as the
so-called wimpzillas,
have been proposed, with masses as large as $10^{15} \mbox{ GeV}$,
i.e. well above the unitarity limit (see e.g. \cite{Wimpzillas
Review} for a review). For mechanisms that produce these
super-massive particles with
$\Omega_{DM}\sim 1,$ departure from thermal equilibrium is automatic
\cite{Wimpzillas Review}, and the challenge is not to overproduce
them. Several mechanisms have been proposed: for instance they could be
created during reheating after inflation, with masses a factor $10^3$
larger than the reheating temperature \cite{Reheating
Wimpzillas}, or during a pre-heating stage, with masses up to the
Grand Unification scale ($10^{15}$ GeV) \cite{Preheating Wimpzillas
GUT} or even to the Planck Scale \cite{Preheating Wimpzillas
Planck}, or again from bubble collisions if the inflation exit is
realized by a first-order transition \cite{Bubble Wimpzillas}.
Another very interesting mechanism is of gravitational nature:
wimpzillas may be created by amplification of quantum fluctuations
in the transition between the inflationary regime and the matter
(radiation) dominated one, due to the nonadiabatic expansion of the
spacetime \cite{Riotto Chung Kolb,Kolb Starobinski}. This scenario
can produce particles with mass of the order of the inflaton
mass and do not require couplings of wimpzillas with inflaton or
other particles.

These particles can be accomodated in existing theoretical frameworks. For
instance, stable or metastable bound states called cryptons arise in
M-theory, and other possibilities are contemplated in string
theories \cite{pyrons}. Furthermore, messenger bosons in soft supersymmetry
breaking models may be very massive and in presence of accidental
symmetries in the messengers sectors, might be stable
\cite{yanagida}.

Although wimpzillas have been invoked in top-down scenarios that
seek to explain the origin of Ultra High Energy Cosmic Rays
\cite{Kolb Starobinski,Sarkar,Ziaeepour}, this interpretation is
today problematic because it predicts a large photon component in
the UEHCRs spectrum, in disagreement with the recent results of the
Auger experiment \cite{Semikoz}.

Several production mechanisms can act together to produce a given
species, and its relic abundance receives contributions from each of
them. The calculation depends on the details of the particle physics
and cosmological models adopted. In the case of axions, i.e. light
pseudoscalar particles introduced to solve the Strong CP Problem,
the production mechanisms in the early Universe are scattering in
the hot thermal plasma and possibly radiation by topological defects
like axion-strings. Another relevant production mechanism is the
so-called {\it misalignment}: the axion field rolls towards its
minimum, near the QCD epoch, and it ends with coherent oscillations
that produce a Cold Dark Matter condensate. A lower bound on the
axion mass can be inferred requiring that they do not overclose the
Universe, but the uncertainties in the calculation of their
production make the constraint rather weak (for recent reviews of
axions see
\cite{Sikivie,Raffelt Axions}).\\

As mentioned before, gravitinos can be copiously emitted by the
decay of the NLP in SUSY but they can also be produced, during
reheating, by inelastic 2$\rightarrow$2 scattering processes
off particles in the thermal bath and in some scenarios they can
act as Cold Dark Matter candidates (e.g. \cite{graviCDM1,
graviCDM2, graviCDM3, Moroi TRehe, graviCDM5, graviCDM6}, see e.g.
\cite{SteffenRevGrav, Buchmuller gravitino Leptog} for more
details and references on gravitino DM models). The efficiency
of the production depends on the reheating temperature $T_R$ so the
bound on $\Omega_{DM}$ translates into an upper limit on $T_R$
\cite{Boltz Brandeburg Bucch, Moroi TRehe, Pradler}. In addition to
thermal production and late decays of the NLSP, other non thermal
and inflation model dependent contributions can arise and change
considerably the predictions \cite{Giudice Riotto,
NillesPelosoSorbo}.

Sterile neutrinos, which arise naturally in theoretical frameworks
\cite{GUT Sterile,Mirror Sterile,String Sterile,ExtraDimension
Sterile} or in the phenomenological $\nu$MSM \cite{nuMSMM}, have
been proposed as a solution of the LSND anomaly \cite{LSND}, as
explanation of the high pulsar velocities \cite{Pulsar velocities}
and as Dark Matter candidates. Recently, the MiniBoone collaboration
has reported its first results, excluding at 98\% C.L. the
two-neutrino appearance oscillation scheme obtained from LSND data
\cite{Aguilar MiniBoone}. The (3+1) scheme, involving one sterile
neutrino specie, is excluded and also models with two or three
sterile species are not viable because of the tension between
appearance and disappearance data \cite{Maltoni Schwtz}.

Sterile neutrinos may be produced in the early Universe from
collision oscillation conversions of active thermal neutrinos. Their
momentum distribution is significantly distorted with respect to a
thermal spectrum due to the effects of quark-hadron transition, the
modification of the neutrino thermal potential caused by the
presence of thermal leptons and the heating of the coupled species
(see e.g. \cite{Abazajan Production} for precise computation of
relic abundance). Moreover it has been proposed an enhanced resonant
production, in presence of a lepton asymmetry in the early universe
significantly higher than the baryonic one \cite{Abazajian
Fuller,Resonant Sterile}.

\section{IS IT COLD?}
\label{sec:chapter two}

The evolution of perturbations in the Universe depends on the
microscopic properties of DM particles. The standard picture, widely
accepted, is that after equality, when the Universe becomes Matter
Dominated, the DM density perturbations begin to grow, and drive the
oscillations of the photon-baryonic fluid around the DM
gravitational potential wells. Soon after recombination, baryons
kinematically decouple from photons and remain trapped in DM
potential wells. Their density perturbations then grow to form the
structures that we observe today in the Universe (see for more
details \cite{Kolb Turner,Dodelson}).

\subsection*{Hot Dark Matter}

The imperfect coupling between baryons
and photons at recombination leads to a damping of small scale
anisotropies, also known as Silk damping \cite{Silk Damping}.
A collisionless species, moving in the universe from higher to lower
density regions, also tends to damp the fluctuations above its
free-streaming scale. This a key property of Hot Dark Matter, which
consists of species which are relativistic at the time of structures
formation and therefore lead to large damping scales \cite{Bond
Szalay}.

The prototype of HDM are Standard Model neutrinos: they were
thermally produced in the early Universe and they termodinamically
decoupled again relativistic at $T \sim 1 \mbox{ MeV}$, leading to a
relic abundance today that depends on the sum of the flavor
masses, $m_{\nu} = \sum_{i=1}^{3} m_{\nu_i} $:

\begin{eqnarray}
\Omega_{\nu}h^2 = \frac{m_{\nu}}{90\mbox{ eV}}.
\label{eqn:Omeganeutrino}
\end{eqnarray}

Their free-streeming length is
\cite{Kolb Turner}:
$$ \lambda_{FS} \sim 20 \left( \frac{30 \mbox{ eV}}{m_{\nu}} \right) \mbox{ Mpc}.$$

Hot DM models are today disfavored (see e.g.
\cite{HOTReview} for a more complete discussion). For instance,
the power spectrum of density perturbations should be suppressed beyond the
free-streaming length of HDM particles, that for neutrino masses in the eV range
corresponds roughly to the size of superclusters.
Furthermore, HDM models predict a top-down hierarchy in the
formation of structures, with small structures forming by
fragmentation of larger ones, while observations show
that galaxies are older than superclusters.

Small amounts of HDM can still be tolerated, provided that it is
compatible with large scale structure and CMB data. Assuming an
adiabatic, scale-invariant and Gaussian power spectrum of primordial
fluctuations, WMAP data set an upper limit on the sum of light
neutrino masses~\cite{Spergel} (or equivalently, through Eq.
\ref{eqn:Omeganeutrino}, on $\Omega_{\nu}$)
$$\sum m_{\nu } < 2.11 \mbox{ eV} \mbox{  (95 \% CL)}.$$
The combination of data from WMAP, large scale structure and
small-scale CMB experiments, further strengthens the constraint, but
it also introduces potentially large systematic effects
\cite{Elgoroy, Spergeletal, Hannestad, Pierpaoli}. A significantly
improved constraint can been obtained combining Ly-$\alpha$ forest,
CMB, SuperNovae and Galaxy Clusters data \cite{Seljak neutrino,
MelchiorriFogli}:
$$\sum m_{\nu } < 0.17 \mbox{ eV} \mbox{  (95 \% CL)}.$$

These limits can be applied to a generic hot Dark Matter candidate,
e.g. to
thermal axions \cite{Raffelt Axions, Raffelt Hannestad, Hannestad Mirizzi Raffelt}
or to hot sterile neutrinos \cite{Dodelson Melchiorri}.\\

\subsection*{Cold Dark Matter}

The standard theory of structure formation thus requires that Dark
Matter is {\it cold}, i.e. it is made of particles that have become
non-relativistic well before the matter domination era, and that can
therefore clump on small scales. The prototype of cold DM candidates
is the supersymmetric neutralino, whose free-streaming length is
such that only fluctuations roughly below the Earth mass scale are
suppressed \cite{CDM freeStreaming, GreenHofmannScwatz}. CDM
candidates can be heavy thermal relics, such as the aforementioned
neutralino, but also light species, non-thermally produced, like
axions (see Sec. \ref{sec:chapter one} for further comments and
references).

N-body simulations of $\Lambda$CDM Universe are in agreement with a
wide range of observations, such as the abundance of clusters at $z
\leq 1$ and the galaxy-galaxy correlation functions (see e.g.
\cite{Primack Review} for a review of CDM), making it a successful
and widely accepted cosmological model.

However, the emergence of some discrepancies has lead some authors
to question the CDM model and to propose alternative scenarios. For
example, the number of satellite halos in Milky Way-sized galaxies,
as predicted by simulations, exceeds the number of observed Dwarf
galaxies \cite{Moore MissingSatellite, Klypin Prada}. Furthermore,
the rotation curves of low surface brightness (LSB) galaxies point
to DM distributions with constant density cores rather than the
cuspy profiles preferred by N-body simulations \cite{Flores
LSS,McGaugh LSS, GentileSalucci, GentileSalucci2}. An additional
problem arises when considering the angular momentum of dark matter
halos: in simulations gas cools at early time into small mass halos,
leading to massive low-angular momentum cores in conflict with the
observed exponential disks \cite{AngularMom Problem}.

Several astrophysical processes have been invoked in order to solve
these problems, such as major mergers and astrophysical
feedback\cite{AngularProbl resolution}. The low efficiency of gas
cooling and star formation may decrease the number of satellites in
Milky Way-sized galaxies \cite{Bullock SatelliteProblem, Gnedin,
MooreDiemandMadau} and tidal stripping may have dramatically reduced
the size of these substructures or disrupted a fraction of them
\cite{MayerGovernatoColpi, KravtsovGnedinKlypin}. Furthermore, new
ultra-faint dwarf galaxies have been recently detected, alleviating
the discrepancy between CDM predictions and observations
\cite{SimonGeha}. It has also been pointed out that the measurements
of the LSB galaxies rotation curves may suffer of observational
biases, for example due to the fact that DM halos are triaxials
rather than spherically symmetric \cite{Navarro}. Moreover, small
deviations of the primordial power spectrum from scale invariance,
the presence of neutrinos \cite{FlatHaloHarrZeld} or astrophysical
processes \cite{El Zant, M Weinberg} can sensibly affect the halo
profiles. Anyway, the lack of convincing explanations of the
problems discussed above leaves the door open to alternatives to the
CDM scenario.

\subsection*{Warm Dark Matter}

To alleviate these problems, Dark Matter candidates with a strong
elastic scattering cross section (SIDM) \cite{SIDM}, or large
annihilation cross sections \cite{SADM} have been proposed. It has
also been suggested that Dark Matter is {\it warm}, i.e. made of
particles with velocity dispersion between that of HDM and CDM
particles. The larger free-streaming length of WDM, with respect to
CDM, reduces the power at small scales, suppressing the formation of
small structures \cite{Bode,Larsen Dolgov}. For instance, a WDM
particle with a mass of 1 keV and an abundance that matches the
correct Dark Matter density, has a free-streaming length of order of
galaxy scales $\lambda_{FS} \sim 0.3 \mbox{ Mpc}$
\cite{MattLeVielgravitino}. Measurements of the growth of structures
in galaxy clusters and Ly-$\alpha$ forest can then be used to set a
lower bound on the mass of the WDM particle. Gravitinos in
gauge-mediated supersymmetry breaking models might be warm DM
candidates, if they decouple when the number of degrees of freedom
was much larger than at the neutrino decoupling \cite{Baltz
gravitino}. However, explicit computations show that such a light
thermal gravitino cannot account for all the DM
\cite{MattLeVielgravitino}.

Another WDM candidate is the sterile neutrino, produced in the early
Universe by oscillation conversion of thermal active neutrinos,
with a momentum distribution significantly suppressed and distorted
from a thermal spectrum \cite{Widrow dodelson,nuMSMM,Abazajian
Fuller}.
Its free-streaming scale is given by (see e.g. \cite{Abazajan Koushiappas})
$$     \lambda_{FS} \thickapprox 840 \mbox{ Kpc}\mbox{ h}^{-1} \left( \frac{1\mbox{ KeV}}{m_{s}}\right) \left( \frac{<p/T> }{3.15}       \right),$$
where $m_s$ is the mass state associated to the sterile flavor
eigenstate. $<p/T>$ is the mean momentum over temperature of the
neutrino distribution and the ratio $<p/T>/3.15$ ranges from
$\thickapprox 1$ for a thermal WDM particle to $\thickapprox 0.9$,
for a non-thermal sterile neutrinos distribution.

The suppression of the power spectrum by a thermal WDM of a given
mass $m_{WDM}$, is identical to that produced by sterile neutrinos
of mass $m_s$ derived by \cite{dodelson colombi,MattLeVielgravitino}:
$$   m_{s}  = 4.43 \mbox{ KeV} \left(\frac{m_{WDM}}{1\mbox{ KeV}}\right)^{4/3} \left( \frac{0.25(0.7)^2}{\Omega_{WDM}} \right)^{1/3}.$$
This one-to-one correspondence allows to translate the
bounds on sterile neutrinos to a generic thermal relic
and viceversa.

A detailed analysis of the production of sterile neutrinos and of
the evolution of their perturbations, as well as a comparison with
the measured matter power spectrum, have been performed in
Refs.~\cite{MattLeVielgravitino, Abazajan2006, Viel2007, Seljack
Sterile, VielBeckerBolton}). The resulting lower limits on the mass
of the WDM particles strongly depend on the dataset used in the
analysis. For example, in \cite{Abazajan2006}, a combination of the
SDSS 3D power-spectrum and SDSS Ly-$\alpha$ forest allowed to
constrain the sterile neutrino mass to
$$m_s \geq 1.7 \mbox{ KeV (95 \%CL)},$$
that translates in terms of a thermal WDM particle to
$$m_{WDM} \geq 0.50 \mbox{ KeV}.$$

The inclusion of high resolution Ly-$\alpha$ data makes the
constraint even stronger, even if it has been pointed out that they may suffer of large
systematic uncertainties \cite{Abazajan2006, MattLeVielgravitino}.

More recently, very stringent bounds on the mass of WDM
particles have been obtained by different groups: \cite{Seljack
Sterile}
$$m_s \geq 14 \mbox{ KeV} \mbox{ (95 \% CL)  } (m_{WDM} \geq 2.5 \mbox{ KeV}) $$
and \cite{VielBeckerBolton}:
$$m_s\geq 28 \mbox{ KeV} \mbox{ (2$\sigma$)  } (m_{WDM} \geq 4 \mbox{ KeV}).$$

The delay of the reionization of the Universe also sets a constraint
on the WDM mass \cite{barkana, YoshidaSokasianHernquist, Jedamzik
Lemoine}. In the case of sterile neutrinos, the X-rays produced by
their decays can modify the picture, enhancing the production of
molecular hydrogen and releasing heat in gas clouds \cite{Bierman,
Stasielak, Ripamonti Ferrara Mapelli}.

\section{IS IT NEUTRAL?}
\label{sec:chapter three}

Some extensions of the Standard Model of particle physics predict
the existence of new, stable, electrically charged particles, such
as the lightest messenger state in gauge-mediated supersymmetry
breaking models \cite{Dimopoulos Giudice Pomarol} or even the LSP in
the R-parity conserving Minimal Supersymmetric Standard Model
(MSSM).

Massive charged particles, independently on the context they emerge,
have been proposed as Dark Matter candidates by De
R$\acute{\mbox{u}}$jula {\emph et al} and dubbed CHAMPs \cite{De
Rujula}. Evaluating their thermal relic abundance, with simple
assumptions on the annihilation cross sections, the authors found a
viable mass range of $\sim 1-1000 \mbox{ TeV}$. They also pointed
out that a positively charged particle $X^+$ can capture an electron
to form a bound state chemically similar to an heavy hydrogen atom.
An $X^-$ can instead bind to an $\alpha^{++}$ particle and an
electron, resulting again in a heavy hydrogen-like atom, or
alternatively it can capture a proton to produce a bound state
called neutralCHAMP. The different behaviors of CHAMPs and
neutralCHAMPs lead to different bounds on their abundance. Note also
that De R$\acute{ \mbox{u}}$jula {\emph et al.} concluded that $X^-$
would emerge from Big Bang Nucleosynthesis preferentially in the
form of neutralCHAMPs \cite{De Rujula}.

Galactogenesis models provide constraints on the Dark Matter
interactions, in particular of CHAMPs. The energy loss timescale in
this case is in fact dominated by Coulomb scattering off protons,
and it must be longer than the dynamical timescale for galaxy
formation. In Ref. \cite{De Rujula}, the
authors concluded that only CHAMPs heavier than $20 \mbox{ TeV}$ are
able to remain suspended in the halo, and to be therefore rare on
Earth. This estimate disagree with that obtained by Dimopoulos
{\emph et al} who found, for the same considerations, the limit $M_X
> 10^5 \mbox{ TeV}$ \cite{Dimopoulos Champs}. It has also been proposed that shock
accelerations in supernovae could eject CHAMPs from the disk and
reinject them back to the halo or out of the galaxy \cite{Dimopoulos
Champs}. The latter possibility is energetically disfavored, while in
the former case, it may lead to a dangerous heating of the disk.

One of the most stringent bounds on the CHAMPs abundance comes from
searches of anomalous heavy water: CHAMPs, being chemically
identical to heavy hydrogen, can be trapped in oceans and lakes in
the form of HXO. If one assumes, as in Ref. \cite{De Rujula}, that
CHAMPs heavier than $20 \mbox{ TeV}$ remain suspended in the
Galactic halo and they provide the Galactic DM, taking an
accumulation time of $3 \times 10^9 \mbox{ yr}$, comparable with the
age of oceans, the abundance of CHAMPs in sea water is predicted to
be \cite{Kudo Yamaguchi}:
$$ \left(
\frac{n_X}{n_H}\right)_{Earth} \sim 3\times 10^{-5}
\left(\frac{\mbox{GeV}}{m_X}\right)\Omega_{X}h^2.$$

If instead CHAMPs are present in the Galactic disk, taking in account the
density and velocity of the interstellar gas, mostly hydrogen, the
expected concentration is \cite{Kudo Yamaguchi}:
$$ \left(
\frac{n_X}{n_H}\right)_{Earth} \sim 6\times 10^{-5}
\left(\frac{\mbox{GeV}}{m_X}\right)\Omega_{X}h^2.$$

All the searches of anomalous hydrogen in sea water have failed
so that the abundance of CHAMPs, for masses in the range
100 GeV-1000 GeV is constrained to be
$$ \left(
\frac{n_X}{n_H}\right)_{Earth} \sim 10^{-28}-10^{-29},$$ while it
raises to $(n_X/n_H)<10^{-20}$ for $M_X \sim 10 \mbox{ TeV}$ (see
\cite{PDG Champs} for a compilation of upper bounds of heavy
hydrogen from sea water searches). As a result, CHAMPs as DM
candidates are ruled out in the mass range $M_{X} \sim 10-10^{4}
\mbox{ GeV}$.

NeutralCHAMPs would preferentially bind on Earth to nuclei to form
anomalous heavy isotopes. Null searches for these elements, covering a variety of
nuclear species, constrain the NeutralCHAMPs abundance to be $<
10^{-20 }- 10^{-16}$ for $M_X \sim 100-1000 \mbox{ GeV}$
\cite{Hammick} (for further details see \cite{PDG
Champs} and references therein). The authors of Ref. \cite{Hammick},
concluded that stable $X^-$ Dark
Matter in the mass range $10^2-10^4 \mbox{ GeV}$ is thus to
be considered unlikely.

CHAMPs are also constrained by balloon or satellites experiments for
Cosmic Rays studies. Perl {\emph et al}, taking in account data from
different experiments \cite{Dimopoulos Champs, Barwick, Snowden},
excluded CHAMPs as Galactic Dark Matter in the mass range
$2.4\times10^{3}-5.6 \times10^{7}\mbox{ GeV}$ and neutralCHAMPs for
$10^{5}-4\times10^{7}\mbox{ GeV}$ \cite{Perl Rev Champs}. The lower
limit comes from the requirement that particles penetrate the solar
wind and the energy deposition is above the experimental threshold.
The upper bound is obtained comparing the maximum CHAMP flux at the
top of the atmosphere allowed by the CR experiments, with the local
DM flux, which is typically assumed to be $\phi \sim 10^7
(\mbox{GeV}/M_X) \mbox{ cm}^{-2}\mbox{s}^{-1}$.

In the atmosphere, a proton in a neutralCHAMP gets replaced very quickly
by a $^{14}N$ atom, and the exchange is followed by a MeV
$\gamma$-ray emission from the excited $^{14}NX^-$ status. With the
same argument explained above, the observational limits on
$\gamma$-rays flux imply that neutralCHAMPs should be heavier than $10^{6}
\mbox{ GeV}$ if they are to be the DM\cite{Dimopoulos
Champs}.
\begin{figure}[tb]
\includegraphics[width=8.5cm]{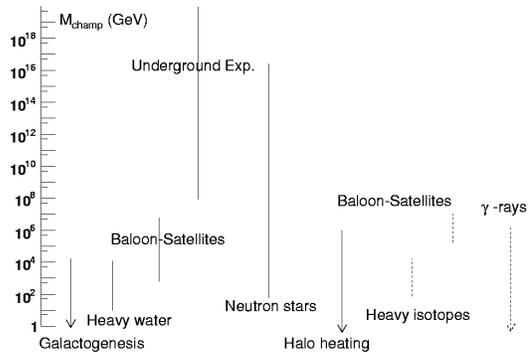}
\caption{Exclusion plot for CHAMPs (solid lines) and NeutralCHAMPs
(dotted lines). See text for more details.} \label{fig:Champs}
\end{figure}
Further constraints on CHAMPs come from deep
underground experiments. The responses of scintillators
to monopoles and CHAMPs are expected to be similar, since they
are both slowly moving, highly ionizing and penetrating. In
Ref. \cite{Perl Rev Champs}, the authors applied the upper limit on
monopole flux, obtained from MACRO experiment, to the CHAMP case,
excluding the mass range $10^{8}-10^{20} \mbox{ GeV}$.

Further constraints come from stellar evolution, in particular it
has been shown that CHAMPs can disrupt a neutron star in a short
timescale, falling into its center and producing a Black Hole. This
argument excludes CHAMPs with masses $10^{2} - 10^{16} \mbox{ GeV}$
\cite{Gould}. In addition, the properties of diffuse interstellar
clouds constrain the interactions of halo particles with atomic
hydrogen: the rate of energy deposition due to collisions must be
smaller than the cooling rate, for clouds in equilibrium. It results
that CHAMPs with masses below $10^{6} \mbox{ GeV}$ are ruled out
because, for these particles, the expected cross section with
hydrogen is higher than the maximum allowed value \cite{Chivukula}.

The various constraints on CHAMPs that we have discussed
are summarized in Fig. \ref{fig:Champs}. Even if the bounds
are not completely model-independent, the combination of them
basically rules out CHAMPs as DM.\\

The above limits apply to particles with integer electric charge,
but theoretical frameworks have been proposed where particles with
fractionary electric charge exist, also known as milli-charged
particles \cite{Abel, Abel Schofield, Holdom, Holdom2, Masso
Redondo, Brian Gherhetta}. For example, adding a new unbroken
$U(1)^{'}$ gauge group, the photon and paraphoton can mix, and
particles charged under $U(1)^{'}$ can have a small coupling with
photons \cite{Holdom}. Moreover, realistic extensions of SM
motivated by string theory exist, that naturally implement this
mechanism \cite{Abel}.

\begin{figure}[tb]
\includegraphics[width=9.0cm]{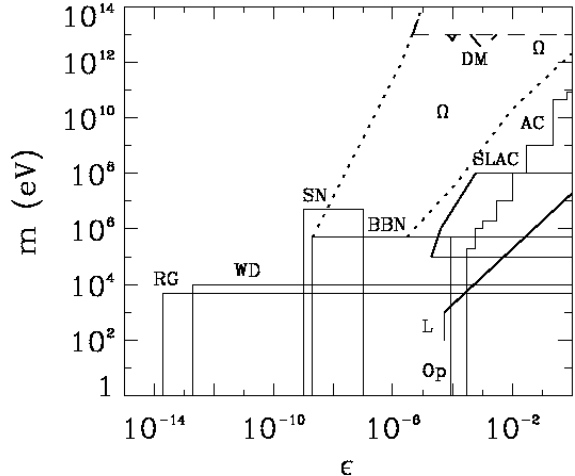}
\caption{Excluded regions in the mass-charge plane for milli-charged
particles. The constraints are relative to: RD plasmon decay in red
giants; WD plasmon decay in white dwarfs; BBN big bang
Nucleosynthesis; SN Supernova 1987A; AC accelerator experiments;
SLAC SLAC millicharged particle search; L Lamb Shift; Op invisible
decay of ortho-positronium; DM Dark Matter searches. From Ref.
\cite{Raffelt Millicharged}.} \label{fig:Millicharged}
\end{figure}

Constraints on mass and charge of milli-charged particles come from
a variety of observations, and in Fig. \ref{fig:Millicharged} we show
the excluded regions in the parameter space ($m_q$, $\epsilon$),
with $\epsilon=q/e$, obtained by Davison {\emph et al.} \cite{Raffelt
Millicharged}.

Milli-charged particles can also affect CMB anisotropies, and for
this reason WMAP data can severely constrain their cosmological
abundance, at least in some regions of the milli-charged particle
parameter space \cite{Dubovsky}.

Furthermore, searches of neutrino magnetic moment with reactor
experiments exclude Dark Matter particles with $q > 10^{-5} e$, for
masses $m_q \lesssim 1 \mbox{ keV}$ \cite{Rubbia Gninenko}.

The result of the PVLAS collaboration \cite{Zavattini} have been
tentatively interpreted in terms of milli-charged particles with
masses $m_q\sim 0.1 \mbox{ eV}$ and fractional electric charge
$\epsilon \sim 10^{-6}$ \cite{Abel, Masso Redondo, Gies}, but the
experimental result was not confirmed after an upgrade of the PVLAS
apparatus \cite{PVLas2007}.

Light milli-charged particles may largely affect sub eV Cosmology.
In particular, processes such as $\gamma \gamma \rightarrow
q\bar{q}$ can distort the CMB energy spectrum, which has been
measured with high sensitivity by FIRAS. A detailed analysis has
been performed in Ref. \cite{Melchiorri} and the authors reported
the conservative upper bound $\epsilon \lesssim 10^{-7}$, for
$m\lesssim 1 \mbox{ eV}$, excluding in this way also the light
milli-charged particles proposed in Ref.
\cite{Abel, Masso Redondo, Gies}.\\

In principle, DM particles could have a $SU(3)_c$ charge. For
example, "colored" candidates are naturally predicted in SUSY models
if the LSP is a squark \cite{Sarid squark} or a gluino \cite{Farrar
Gluino, Rabi Gluino}, or in gauge mediated SUSY breaking models,
where messengers can be colored and stable \cite{Chacko Messenger
Colored}, or in mirror models \cite{Beherenziani Mirror Colors}.
These "heavy partons", after the deconfinement temperature, $T\sim
180 \mbox{ MeV}$, are surrounded by a QCD cloud and confined inside
hadrons forming a color neutral bound state \cite{Kang Nasri}. These
particles can be actively searched for by underground experiments,
indirect detection experiments or through the search of rare
anomalous isotopes.

Since the proposal that DM might interact strongly with ordinary
matter (SIMP), regardless of the nature of the interaction \cite{De
Rujula, Dimopoulos Champs, Starkman}, many
candidates have been put forward, but also many constraints on the
scattering cross section off nuclei, $\sigma_{\chi N}$.

For example, the SIMPs interactions with baryons may disrupt the
disk of spiral galaxies \cite{Starkman, Natarajan Loeb}. Moreover,
they may dissociate the light elements produced during Big Bang
Nucleosynthesis, while SIMPs collisions with Cosmic Rays can produce
an observable $\gamma$-ray flux \cite{Cyburt SIMP}. The
scattering of SIMPs off baryons also produces substantial distortion
of CMB anisotropies and of the large scale structure power
spectrum \cite{Chen Hannestad}. The SIMPs abundance for the mass
range $ \sim 1-10^3 \mbox{ GeV},$ is also constrained by searches in
terrestrial samples of gold and iron \cite{Javorsek}.

Atmospheric and satellite experiments, originally intended for other
purposes, have been used to investigate high DM cross sections with
baryonic matter. In particular, the results of the X-ray Quantum
Calorimeter experiment (XQC) allow to rule out a large portion of
the SIMP parameter space ($M_{\chi}$, $\sigma_{\chi N}$), as
discussed in Refs. \cite{Wandelt, McGuire Steinharedt} and (more
recently and with substantial changes with respect to previous
analyses) in Ref. \cite{Erickckec Steinhradt}.

Complementary constraints are obtained by underground experiments,
which are sensitive to DM particles with small interactions. In fact,
they are able to detect SIDM particles if their interactions with
ordinary matter are high enough to trigger a nuclear recoil in the
detector but at the same time low enough to allow the particles to
penetrate the Earth crust to the detector \cite{Albouquerque}.
\begin{figure}[tb]
\includegraphics[width=8.5cm]{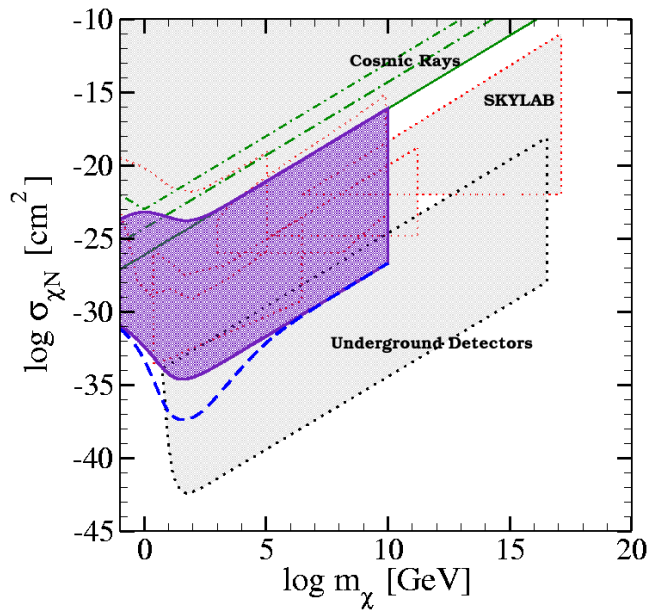}
\caption{Excluded regions in the SIMP mass versus SIMP-nucleon cross
section plane. The Violet area is excluded by the Earth's heat
argument. See Ref.\cite{Mack Beacom Bertone SIMP} and references
therein.} \label{fig:Mack}
\end{figure}
Recently, Mack {\emph et al.} have analyzed the effect of SIMP
annihilations on Earth, showing that a substantial heating of the
Earth's core may occur, if the capture rate is efficient \cite{Mack
Beacom Bertone SIMP}. This argument rules out the regions of the
parameter space lying between astrophysical and underground detector
constraints.

To summarize the constraints on the SIMP scenario, Fig.
\ref{fig:Mack} shows the excluded areas in the SIMP parameter space.
The bounds leave no room for SIMPs as Dark Matter candidates in a
very large mass range. Since the neutron-neutron scattering cross
section is of order $10^{-25}-10^{-24} \mbox{ cm}^2$ and the
expected value for colored Dark Matter candidates is not far from
this range ( see e.g. \cite{Nussinov}), DM particles are thus
unlikely to bring color charge.

However, these constraints can be evaded by very massive composite
dark matter candidates. For example macroscopically large nuggets of
ordinary light quarks and/or antiquarks, with masses in the range $m
\sim 10^{20}-10^{33} \mbox{ GeV},$ can behave as collisionless cold
dark matter, without contradicting observations \cite{Zhitnitsky}.

\section{IS IT CONSISTENT WITH BBN?}
\label{sec:chapter four}

Big Bang Nucleosynthesis (BBN) is one of the most impressive
successes of the Big Bang Cosmology (See \cite{Sarkar? vedi PDG,
Olive Review} for reviews). It predicts the abundances of light
elements produced in the first 3 minutes after the Big Bang, in
agreement with the observations over a range spanning nine
order of magnitudes.

The model is based on a set of coupled Boltzmann equations relating the
number densities of protons, neutrons and light elements, through a
network of nuclear chemical reactions. The weak interactions
maintain the neutron-proton ratio to its equilibrium value until the
freeze-out, that occurs at roughly 0.7 MeV. Later, nearly all
neutrons are captured in the nuclei producing principally the most
stable element $^4$He. Smaller amount of $^2$H, $^3$H, and $^7$Li
are synthesized but the production of heavier elements is suppressed
by the large Coulomb barriers.
Looking for astrophysical environments with low metallicity, it is
possible to infer the primordial abundance of light elements in
order to test the predictions of BBN.

In the framework of the Standard Model, BBN depends only on the
baryon to photon ratio $\eta$, and observations of the abundance of
different elements agree with predictions in the range \cite{Sarkar?
vedi PDG}
$$ 4.7 \leq \eta \cdot10^{10} \leq 6.5 \mbox{ (95\% CL)}.$$
The agreement between predictions and measurements is a powerful
success of the model and it is remarkable that the inferred
abundance of baryons quoted above is also consistent with the
estimate of CMB experiments like WMAP \cite{Spergel}. BBN also
provides a test of physics beyond the Standard Model, and it also
constraints deviations from Standard Cosmology. In fact, the
primordial abundance of $^4$He is proportional to the ratio n/p and
its value is related to the freeze-out temperature of the weak
interactions and it is therefore sensitive to the expansion rate at
that time.

Since $H \propto g_*^{1/2} T^2$, an increase of the relativistic
degrees of freedom $g_*$ with respect to the SM value leads to a
faster expansion rate, thus to an earlier freeze out of the neutron to
proton ratio and consequently to a higher $^4$He abundance (and in
general it also affects the abundance of the other light elements).
At T$\sim1$ MeV, the relativistic species in the Standard Model are
photons, electrons and neutrinos so with $N_{\nu}$ neutrino family
$g_* = 5.5 + \frac{7}{4} N_{\nu}$ and for $N_{\nu}=3$ this gives
43/4.

New relativistic particles can be accounted for through the
introduction of an effective number of neutrinos:

$$ \frac{7}{4} (N_{\nu} -3) =
  \sum_{i=\mbox{extra b}} g_i
\left(\frac{T_i}{T}\right)^4 + \frac{7}{8} \sum_{i=\mbox{extra  f}}
g_{i} \left(\frac{T_i}{T}\right)^4  \nonumber,
$$

where $T_i$ parametrizes the energy density of the relativistic species and
b (f) stands for bosons (fermions).

A likelihood analysis, taking $\eta$ and $N_{\nu}$ as free parameters,
and based on the abundance of $^4$He and $^2$H, constrains the
effective number of neutrinos to be \cite{Cyburt Nnu}:
$$1.8<N_{\nu}<4.5 \mbox{ (95\% CL)}.$$
Assuming the value of $\eta$ inferred by CMB experiments,
the limit is further strengthened to \cite{Cyburt Nnu}:
$$ 2.2 < N_{\nu} <4.4 \mbox{ (95\% CL)}.$$

These bounds on $N_{\nu}$ can be applied to new species affecting
the expansion rate during nucleosynthesis, such as gravitons
\cite{Maggiore}, neutrinos with only right-handed interactions
\cite{Cyburt Nnu} or millicharged particles \cite{Raffelt
Millicharged}. For a large class of supergravity models with a light
gravitino, the requirement $N_{\nu}<4$ rules out gravitino masses
below 1 eV \cite{Grifols}. However, particles coupled to photons or
to neutrinos during BBN, with masses in MeV range, have a non
trivial impact on BBN, that cannot be accounted for with an
equivalent number of light neutrinos \cite{Kolb Turner Walker}.

For instance, it has been suggested that MeV Dark Matter, with masses
in the range 4-10 MeV and coupled with the electromagnetic plasma,
can lower the helium and deuterium abundances, contrary to what one
naively expects, and it can therefore improve the agreement
between the predicted and measured $^4$He abundance \cite{Raffelt
MeVDM}.

In addition, the predictions of BBN can be dangerously modified by
the decays of particles during or after BBN. For example, radiative
dacays induce electromagnetic showers and the subsequent
photon-photon processes can destroy the light elements. In the early
stages of BBN, $t<10^2$ sec, hadronic decays may modify the
interconversion of protons and neutrons, increasing the n/p ratio
and consequently enhancing the $^4$He and $^2$H abundance. The
opposite effect occur for late hadronic dacays, $t>10^2$ sec, when
the energetic hadrons trigger the $^4$He dissociation.

Accurate calculations, with the use of BBN codes, restrict the
primordial abundance of the decaying particle, depending on its
lifetime, mass and hadronic branching ratio \cite{Jedamzik2004,
Kawasaki Kohri Moroi,Steffen,Cybert Olive 2003,Jedamzik2006}. These
results can be applied for instance to NLSP gravitinos: BBN requires
an upper limit to the reheating temperature, which controls the
primordial gravitino abundance, and in some cases the restrictions
could lead troubles to thermal leptogenesis and inflation models
\cite{Kawasaki Kohri Moroi, Cybert Olive 2003}. The difficulties can
be circumvented for example in the case of a heavy gravitino, which
decays well before BBN \cite{Gherghetta,Ibe}.

Alternatively, the gravitino could be stable and play the role of
Dark Matter. In this case, the NLSP particle is typically
long-lived, because of the extremely weak interactions of gravitino,
and its late decays can affect BBN. Moreover, it has been pointed
out that if the long-lived particles are charged, e.g the stau, they
can form bound states with light elements, potentially overproducing
$^6$Li and $^7$Li \cite{Pospelov, Kohri Takayama, Kawasaki Kohri
Moroi 2007, Cyburt Olive 2006, Jedamzik}. However, these elements
can also be destroyed, alleviating the severe bounds on the CHAMPs
abundance during BBN.

A neutralino NLSP is excluded \cite{Feng Su Takayama, Covi
Buchmuller, Cerdeno Choi Jedamzik06}, while sneutrino NLSP poorly
affects BBN \cite{KanzakiKawasakiSneutr}. A stop NLSP is
viable in some regions of the parameter space \cite{DiazCruzStop}.

We note that these BBN bounds can be circumvented if the NLSP abundance
is diluted due to a significant entropy production \cite{HamaguchiHatsuda}.

\section{DOES IT LEAVE STELLAR EVOLUTION UNCHANGED?}
\label{sec:chapter five}

Stellar evolution provides a powerful tool to constrain particle
physics, providing bounds that are often complementary to those
arising from accelerator, direct and indirect Dark Matter searches.

If Weakly interacting particles are light, they may be produced in the hot plasma
in the interior of stars, and if they escape without further interactions, they
represent an energy loss channel for the star, possibly modifying the stellar
evolution. Such particles may also be detected on
Earth, as was the case for neutrinos from SN 1987A, or they can
be indirectly searched for through their decay products. Here we describe the most
important observational consequences (see Refs.
\cite{Raffelt Rev Stars, Raffelt Book} for extensive reviews).

Stars as the Sun can be described as self-gravitating gas in
hydrostatic equilibrium, such that the gas pressure equilibrates the
gravitational force. A significant energy loss produces a
contraction of the system and an increase of the burning rate of the
stellar fuel, reducing the lifetime of the star and enhancing
the neutrino flux. Moreover, exotic energy losses would modify the
sound speed profile, which is accurately measured in the interior of
the Sun by means of helioseismic measurements.

Globular Clusters are alternative interesting probes of stellar evolution models
because they are
gravitationally bound systems of up to a million stars, formed at the
same time, with the same chemical composition and differing
only for their masses. The ignition of helium in Red
Giant stars is sensitive to the temperature and density of the
helium core, and any energy loss channel inevitably tends to delay it, resulting in
more massive cores and producing observational consequences, such as
an enhancement of star brightness. Therefore, observations of Red
Giants in Globular Clusters allow to derive an upper limit on the
energy loss rate of the helium plasma, $\epsilon$, \cite{Raffelt Rev
Stars}:
$$\epsilon \lesssim 10 \mbox{ erg} \mbox{ g}^{-1}\mbox{s}^{-1} \mbox{   at T}\approx 10^{8}\mbox{ K, } \rho\approx2\times 10^5 \mbox{g cm}^{-3},$$
where the value of temperature and density are appropriate for Red
Giant cores. In Horizontal Branch stars, energy losses speed up the
helium burning rate, decreasing their lifetimes, that can be
measured by number counting in Globular Clusters. This argument
provides another bound on $\epsilon$ \cite{Raffelt Rev Stars}:
$$\epsilon \lesssim 10 \mbox{ erg} \mbox{ g}^{-1}\mbox{s}^{-1} \mbox{   at T}\approx 0.7 \times10^{8}\mbox{ K, } \rho\approx0.6\times 10^4 \mbox{g cm}^{-3}.$$

In addition, the cooling rate of White Dwarfs, inferred by their
luminosity functions, is in agreement with the predictions and
therefore any new cooling channel has to be subdominant.


It is remarkable that the total number of neutrino detected from SN
1987A, their energy and their time distribution, is in agreement
with expectations from the standard model which describes the core
collapse of a star. Any further energy loss mechanism reduces
the duration of the neutrino burst and can in principle spoil the
success of the model, leading therefore to the following bound on
$\epsilon$ \cite{Raffelt Rev Stars}:
$$\epsilon \lesssim 10^{19} \mbox{ erg} \mbox{ g}^{-1}\mbox{s}^{-1} \mbox{   at T}= 30\mbox{ MeV, } \rho=3\times 10^{14} \mbox{g cm}^{-3}.$$

All the arguments listed above provide upper limits to any
additional energy loss rate and can be applied to constrain, for
instance, the neutrino properties, the graviton emission in theories
with extra dimensions, as well as models with right-handed
neutrinos, sterile neutrinos, milli-charged particles,   axions and
other pseudoscalar particles. For instance, updated limits on axions
from stars are reviewed in Ref.~\cite{Raffelt Axions Stars} and the
implications of light Dark Matter or sterile neutrino Dark Matter on
Supernovae core collapse are discussed in Ref.~\cite{Fayet SN,
Hidaka SN}. More details and references for other particle physics
scenarios can be found in \cite{Raffelt Rev Stars, Raffelt Book}.

As we have seen in Sec.\ref{sec:chapter three}, the most restrictive
bounds on the fractional charge of $\sim$ keV milli-charged
particles come from stellar physics, as it was shown in
Fig.\ref{fig:Millicharged}.

The bounds discussed above apply to particles that are produced in the core
of stars and that escape without losing energy, thanks to their weak
interactions. However,
if the particles interact strongly, they undergo multiple scattering,
providing a mechanism for energy transport, in competition with photons,
electrons or convection. This effect has been studied for keV-mass
scalars produced in the Sun, Horizontal Branch stars and Red Giants,
constraining the interactions of these particles \cite{Carlson
Scalar, Raff Starkman KeV}.

Moreover, the energy transport channel, provided by the WIMPs
trapped in the Sun, may cool its interior and decrease the neutrino
flux. This idea was proposed in the past as a solution of the solar
neutrino problem and WIMPs with masses and cross sections suitable
for this purpose ($m \sim 4-10 \mbox{ GeV } \sigma \sim10^{-36}
\mbox{ cm}^{-2}$) were called cosmions \cite{Steigman et al, Spergel
Press, Faulkner Gilliland}.

Stars in which the heat transport is dominated by core convection
may be dramatically affected by WIMPs in the case of effective
transport of energy. In fact, in this case,
the convection is suppressed, and the
core is not replenished with nuclear fuel from outer regions,
leading to a reduced stellar lifetime and a modification of its
evolution \cite{Renzini, Bouquet}. As a consequence, main sequence
stars would present an anomalous mass-to-luminosity relation and
Horizontal Branch stars would develop thermal pulses \cite{Bouquet, HB
Deabourn, HB}.
However, taking in account the current limits on the WIMP-nucleon
cross section inferred from direct searches, these effects seems to
be hardly detectable for the Sun \cite{BottinoFiorentini,LopesBertone}.

Dark Matter annihilations may provide an important source of
energy, which, for stars orbiting in high Dark Matter density
regions, can even be comparable or overwhelm that originated by
nuclear reactions. This scenario has been investigated, in the case
of main sequence stars, in Ref.\cite{SalatiSilk} and more recently,
by means of numerical simulations, in Ref.\cite{ScottFairbairn,
ScottFairbairn2}. The most pronounced effects are on low-mass stars
and the authors in Ref.\cite{ScottFairbairn, ScottFairbairn2} have
proposed that these "WIMP burners" could be found in regions where a
recent star formation is inhibited, looking for populations of stars
appearing oddly younger then higher mass ones. Although
apparently young stars have been detected in the inner regions of
Andromeda and of our own galaxy, it is difficult to
interpret these observations in terms of WIMP burners.
WIMP annihilations could also enhance the luminosity of White Dwarfs
and neutron stars, as it has been observed in \cite{Moska Wai, Moska
Wai2, BertoneFairbairn}.

The effect of DM decays and annihilations on the formation of first
structures have been investigated for light DM candidates
\cite{Ripamonti Ferrara Mapelli, Chen Kamionkowski}. Recently, it
has been pointed out that even standard DM candidates, such as the
neutralino, may substantially modify the evolution of Population III
stars, which may even be supported by DM annihilations rather than
nuclear reactions, during part of their evolution~\cite{Gondolo
Freese}.

\section{IS IT COMPATIBLE WITH CONSTRAINTS ON SELF-INTERACTIONS?}
\label{sec:chapter six}

The collisionless and cold nature of Dark Matter, has been questioned
during the last decade, because of apparent discrepancies between the
results of CDM simulations and observations. Two remarkable problems,
as mentioned in Sec. \ref{sec:chapter two}, are the conflict between the cuspy DM
halos predicted by N-body simulations and the constant core profiles
inferred by LSB and dwarfs \cite{Flores LSS,McGaugh LSS} and the
excess of substructures in CDM halo with respect to the observed number
of galaxy satellites \cite{Moore MissingSatellite, Klypin
Prada}.

Although astrophysical explanations exist for the observed
discrepancies, \cite{Bullock SatelliteProblem, Gnedin, Navarro,
FlatHaloHarrZeld, El Zant, M Weinberg}, many attempts have been made
to modify the properties of DM particles in order to reproduce the
appropriate astrophysical phenomenology. A possible solution is that
DM is {\it warm} rather than {\it cold}, as mentioned before.
Alternatively, DM might be self-interacting (SIDM), as proposed by
Spergel and Steinhardt, with large scattering cross section (and
small enough annihilations cross sections, in order to be consistent
with the bounds from indirect detection) \cite{SIDM}. The net
effect, under these assumptions, is that the central cusp reduces to
an almost constant core. Moreover, subhalos can be destroyed by
interactions with the surrounding halo particles, because they are
excessively heated or because particles are scattered out of the
them \cite{SIDM, Wandelt}. Suitable SIDM candidates include Q-balls
\cite{SIDM, Kusenko Shapo Qballs}, a quark-gluino bound state
\cite{Farrar SIDM, Wandelt} and scalar gauge singlets coupled with
Higgs field \cite{Bento Bortolami}.

Semi-analytical calculations and N-body simulations have been
developed to study the effect of SIDM interactions on halo
structures, especially for what concerns the formation of flat cores. Different
solutions are obtained, depending on the ratio between the mean free
path of the Dark Matter particle ($\lambda_{mfp} \propto (\rho
\mbox{ }\sigma/m)^{-1}$, with $n$ the number density and $\sigma /m$
the scattering cross section per unit mass of the SIDM) and the
virial radius of the halo. A cross
section per unit mass in the range $\sigma/m \sim 0.5 -5
\mbox{cm}^2\mbox{g}^{-1}$ was found to correctly reproduce the
observed profile of galaxies \cite{Yoshida, Dave, Colin, Ahn
Shapiro2002}. More recently, Ahn and Shapiro have found a
much higher value, $\sigma/m \simeq 200 \mbox{cm}^2\mbox{g}^{-1},$
as the best fit to LSB rotation curves~\cite{Ahn Shapiro
2005}.

Several constraints exist on SIDM interactions. For instance,
Gnedin and Ostriker have shown that,  for $0.3 \lesssim \sigma / m
\lesssim 10^{4} \mbox{ cm}^2\mbox{g}^{-1},$ galactic halos in
clusters would evaporate in a timescale shorter than an Hubble time
\cite{Gnedin Ostriker}. Following the suggestion of Furlanetto and
Loeb \cite{Furlanetto}, Natarajan {\emph et al.} compared the
predicted truncation radii of SIDM halos with those of observed
galactic halos in clusters, inferred by gravitational lensing,
excluding $\sigma/m
> 42 \mbox{ cm}^2\mbox{g}^{-1}$ \cite{Natarajan}.

An upper limit of $\sigma/m < 0.1 \mbox{ cm}^2\mbox{g}^{-1}$, has
been obtained by Arabadjis {\emph et al.} comparing the results
of simulations with the profile of the cluster MS 1358+62
\cite{Arabadjis}. Hennawi and Ostriker ruled out $\sigma/m \gg
0.02\mbox{ cm}^2\mbox{g}^{-1}$, pointing out that the supermassive
black holes at the center of galaxies would be more massive than
observed \cite{hennawi Ostriker}. The evidence of ellipticity in DM
halos has been used to rule out $\sigma/m >0.02 \mbox{
cm}^2\mbox{g}^{-1}$ because self-interactions tend to produce more
spherical halos \cite{Miralda escude}.
The limits reported above rule out the range of cross
sections required to explain the mass profiles of galaxies, although
the underlying simpliying assumptions
and incomplete statistics suggest to take them with a grain of salt (see e.g.
\cite{Ahn Shapiro 2005, Markevitch}).

More robust results are obtained from the analysis of the 1E 0657-56
cluster of galaxies~\cite{Tucker}, which actually consists of a
bullet-like gas sub-cluster, exiting the core of the main cluster at
high velocity, $v \sim 4700 \mbox{ Km }\mbox{s}^{-1}.$ The
combination of optical and X-ray images with the weak lensing map,
shows that the centroid of the collisionless subcluster galaxies is
ahead of the subcluster gas distribution and coincident with that of
the Dark Matter clump. This cluster not only provides a robust
visual evidence for Dark Matter, but it is also provides a probe of
its collisionless nature.

The subcluster DM halo would be dragged by
the main halo in presence of DM self-interactions, leadig to an
offset between the galaxies centroid and the total mass peak inferred
through weak lensing measurents. Moreover, the measured high merger
velocity implies that eventual
drag forces, due to DM collisions, are small.
Finally, the mass to light ratio of the subcluster is in agreement
with that observed in other clusters and in the main cluster, while SIDM
would tend to scatter out the particles from the subcluster.

Markevitch {\emph et al.} found that the latter argument provides
the most restrictive limit to the self interaction cross section,
$\sigma/m < 1 \mbox{ cm}^2\mbox{g}^{-1}$ \cite{Markevitch}. Making
use of more recent observations and more accurate N-body simulations,
this bound has been slightly improved, $\sigma/m < 0.7 \mbox{
cm}^2\mbox{g}^{-1}$ \cite{Randall Markevitch}. Since this constraint
assumes an identical mass-to light ratio of cluster and subcluster
before the merger, a more robust limit is inferred by the absence of
an offset between galaxies and total mass peaks, which implies
$\sigma/m < 1.25 \mbox{ cm}^2\mbox{g}^{-1}$ \cite{Randall
Markevitch}. Almost the full range of cross sections needed to solve
the discrepancies emerged in CDM models, $\sigma/m \sim 0.5 -5
\mbox{cm}^2\mbox{g}^{-1}$, is ruled out, thus disfavoring
SIDM.

It has been suggested that the scattering
cross section might be velocity dependent, thus
smaller on average in clusters than in galaxies (e.g. \cite{Colin,
Gnedin Ostriker, hennawi Ostriker, Firmani}). A possible functional form is
$$\sigma = \sigma_{*} \left( \frac{100 \mbox{ Km s}^{-1} }{v_{rel}} \right)^{a}.$$
In order to avoid a fast evaporation of the cluster or a core
collapse, the parameters are restricted to be: $\sigma_* = 0.5-1
\mbox{ cm}^2\mbox{g}^{-1}$ and $a =0.5-1$ \cite{Gnedin Ostriker,
hennawi Ostriker}, and simulations have confirmed that in this
range, predictions match the observed flat cores \cite{Colin}.
However, observations of the LSB galaxy NGC 5963 seem to require an
effective cross section per unit mass $ \sigma/m <0.2
\mbox{cm}^2\mbox{g}^{-1},$ in the low velocity regime $\sim 150 \mbox{
km s}^{-1},$  at odds with the quoted range \cite{salcedo}.

In addition to the bounds presented above, that constrain the cross
section per unit mass, additional limits come from the unitary of
the scattering matrix \cite{Unitarity bound, Hui Ubound}. This
argument provides upper bounds on the total cross section and
inelastic cross section. In the low energy regime, taking in account the range of SIDM
elastic cross sections needed to the reproduce the observed halo
profile, the constraint on the total cross section can be turned
into an upper bound on the SIDM mass: $m< 12$ GeV \cite{Hui Ubound}.
Possible exceptions to unitarity bounds have been discussed in
Refs.~\cite{Unitarity bound, Hui Ubound}.

Self-annihilating DM has also been proposed to solve the cold dark
matter cusp crisis \cite{SADM} This scenario is however ruled out by
the comparison of the neutrino flux from the Galactic center with
the measured rate of (atmospheric) neutrinos, i.e. the least
detectable among the final states produced in DM annihilations
\cite{Beacom Bell Mack06}. For the mass range $10^{-1}-10^{5}$ GeV,
$\langle \sigma_{ann} v \rangle$ has to be less than roughly
$10^{-21} \mbox{ cm}^3\mbox{s}^{-1}.$

\section{IS IT CONSISTENT WITH DIRECT DM SEARCHES?}
\label{sec:chapter seven}

Direct DM searches aim at detecting DM particles through the
measurement of nuclear recoils produced by DM scattering. Despite
the large DM flux expected at Earth, $\Phi \sim 10^5 (100 \mbox{
GeV} /m_{DM}) \mbox{ cm}^{-2}\mbox{s}^{-1}$ assuming a local density
of $\rho_0 \sim 0.3 \mbox{ GeV}\mbox{cm}^{-3}$ and mean velocity of
$\bar{v}\sim 220 \mbox{ km}\mbox{ s}^{-1}$, the weakness of WIMP
interactions with nuclei makes direct detection challenging (see
e.g. \cite{Munoz} for a review of direct searches).

The coupling between a WIMP and a nucleon receives contributions
from both scalar (spin independent) and vector
(spin dependent) interactions. The cross section for spin
independent (SI) coupling  with a nucleus (N) cross section
is coherently enhanced with respect to the case of single nucleons:
$$\sigma^{SI}_{N} \simeq A^2(M_{red}(N,M_{\chi})/M_{red}(p,N_{\chi}))^2
\sigma^{SI}_p$$ where $A$ is the atomic number and $M_{red}$ is the
reduced mass of the system WIMP ($\chi$) - nucleus (proton)
\cite{Kurylov Kamion CS}.

Although heavy nuclei are used in current DM direct detection
experiments, the results are often given in terms of scattering
cross section off protons in order to allow easy comparison between
different experimental settings, involving different target
materials. For spin dependent (SD) couplings, there is no coherent
enhancement, and the cross section is determined by unpaired
neutrons or protons in the target nucleus. For this reason, SI
interactions usually dominate the cross section in current
experiments, which exploit heavy nuclei. However, in region of
parameter space where scalar coupling is suppressed, spin dependent
couplings may represent the leading contribution to the direct
detection event rate \cite{Bednya Simkov, Akerib SD}.

The signature of DM
elastic scattering off nuclei are nuclear recoils, characterised by
an exponential recoil spectrum  with typical energies of $\cal{O}$$(10)$ keV
or less, for WIMP masses between 1 and 100 GeV (see for more
details e.g. \cite{Jungman Kamionkowki Griest}). In the case of
inelastic scattering off nuclei
or orbital electrons, the recoil is followed by a decay photon from
the excited state \cite{Ellis Inelastic, Starkman Inelastic}.
However, the natural radioactivity background makes the detection
of this signal very problematic.
\begin{figure}[tb]
\includegraphics[width=8.5cm]{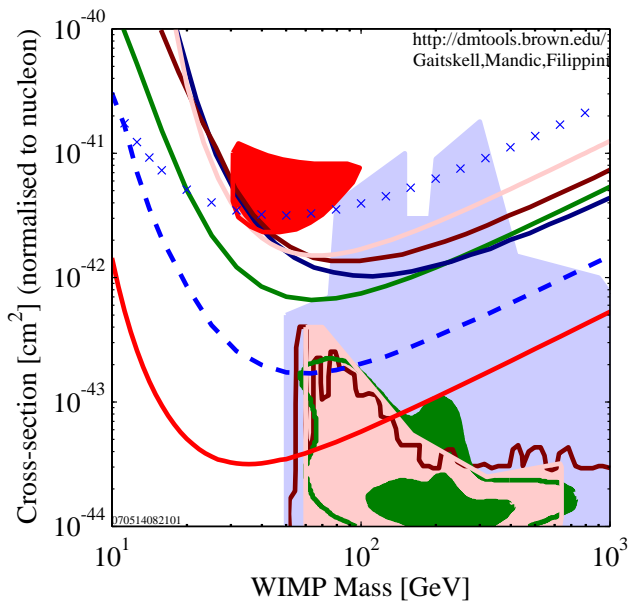}
\caption{Upper limits on the spin independent WIMP-nucleon cross
section, versus WIMP mass. The blue dashed (points) line is the Ge
(Si) CDMS bound \cite{CDMS2006}. The dark red, pink, green and dark
blue curves are the experimental limits respectively from EDELWEISS
\cite{SI EDELW}, CRESST 2004 \cite{SI CRESST}, ZEPLIN II (Jan. 2007)
\cite{SI ZEPLIN} and WARP \cite{SI WARP}. The lowest red solid line
shows the first results from XENON 10 \cite{SI XENON}. The red
shaded region is the parameter space favored by DAMA experiment
\cite{Bernabei LargeA}. Supersymmetric models allow the filled
regions colored: pink \cite{SI Baer }, green \cite{SI Ruiz}, dark
red \cite{SI Baltz} and blue \cite{SI Baltz Gon}. This figure has
been obtained with the use of the interface at
http://dendera.berkeley.edu/plotter/entryform.html.}
\label{fig:SpinIndependent}
\end{figure}
Current experiments exploit a variety of detection techniques,
focusing on signals such as
scintillation, phonons, ionization or a combination
of them, as well as a variety of targets, e.g.
NaI, Ge, Si and Xe.

In order to discriminate a DM signal against the natural background,
some experiments have been searching for an annual modulation of the
measured event rate \cite{Drukier Modulation}. In fact, the Earth
rotation around the Sun is expected to produce a modulation of the
relative velocity of DM particles given by
$$ v_E = 220\mbox{ Km/s } \{ 1.05 + 0.07 \cos
[2\pi(t-t_m)]/1\mbox{ year} \}$$ where $t_m$ is approximatively the
begin of June. The variation of the WIMP flux is actually small $\approx 7
\%$, so that a large number of events has to be collected and therefore a
large detector is needed.

\begin{figure*}[tb]
\includegraphics[width=8.5cm]{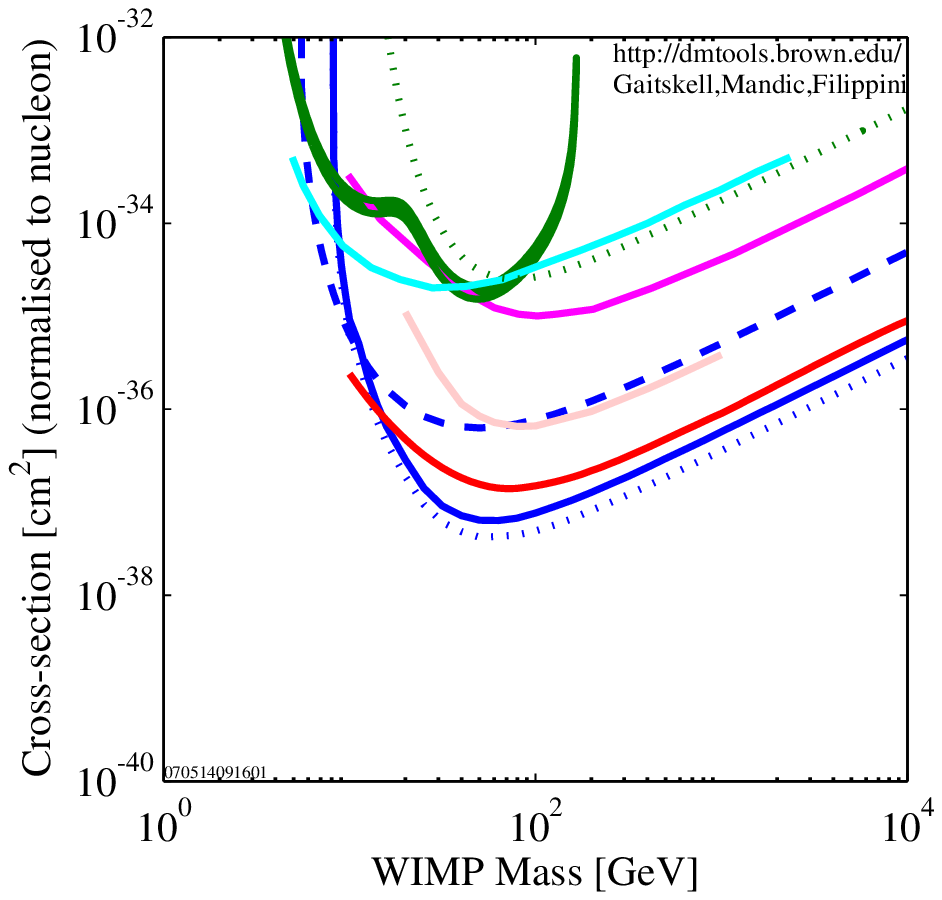}
\includegraphics[width=8.5cm]{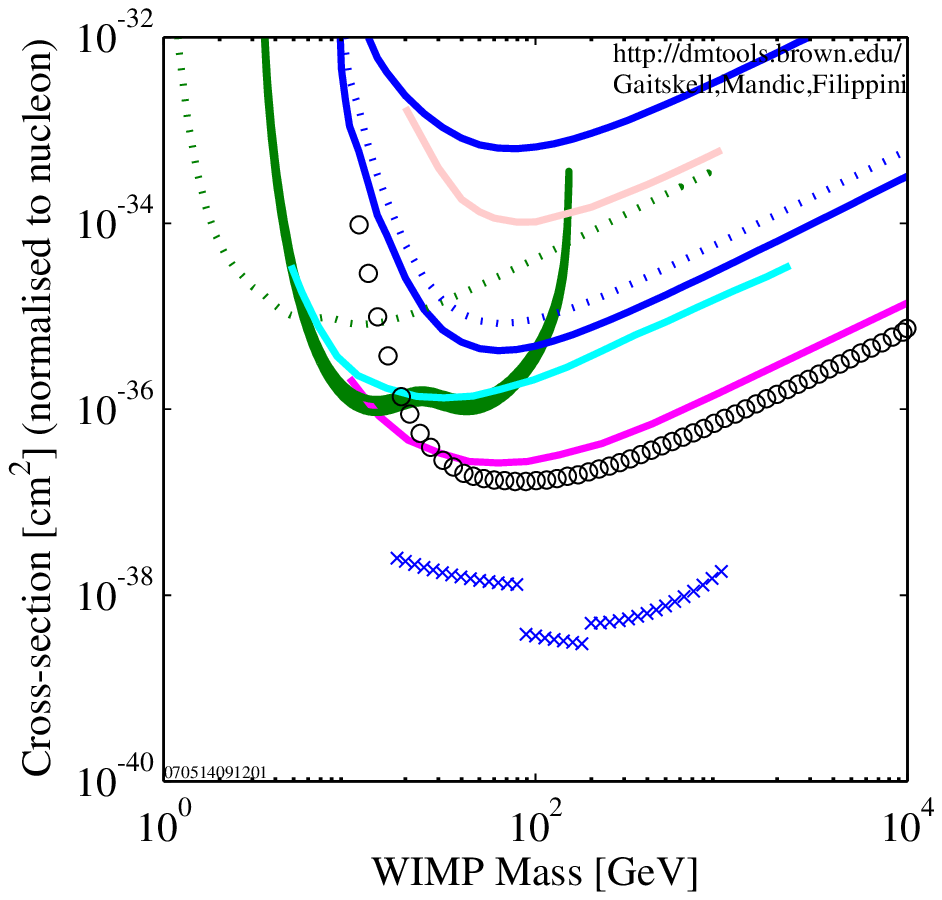}
\caption{Upper limits on spin-dependent WIMP cross section as a
function of the WIMP mass, in the case of a pure neutron (proton)
and proton (left) coupling. The blue solid (dashed) line is the Ge
(Si) CDMS bound \cite{Akerib SD}. The blue dotted line is the CDMS
limit with an alternative form factor \cite{Akerib SD}. The light
red, cyan, magenta and red curves are the experimental limits
respectively from EDELWEISS \cite{SD Edelweiss}, PICASSO \cite{SD
Picasso}, NAIAD 2005 \cite{SD NAIAd} and ZEPLIN I \cite{SD Zeplin}.
The dark green shaded region shows the parameter space favored by
DAMA experiments \cite{SD Cresst}. Finally the green points
represent the CRESST results \cite{SD Cresst}, the black crosses
stand for Super-Kamiokande \cite{SD SuperKAM} and the black circles
for KIMS 2007 \cite{SD KIMs}. The figures have been obtained with
the use of the interface at
http://dendera.berkeley.edu/plotter/entryform.html.}
\label{fig:SpinDedependent}
\end{figure*}

In 1998, the DAMA collaboration obtained evidence for a modulation
of the event rate, that was later confirmed with a confidence level
of 6.3 $\sigma$ (see \cite{DAMA 2006} for a recent discussion).

If interpreted in terms of a SI scattering of a WIMP off NaI,
and further assuming an isothermal sphere DM halo, with a characteristic
velocity of the Maxwell-Boltzmann distribution of $v_0 = 270 \mbox{
Km}\mbox{s}^{-1}$, with a local DM density of $\rho = 0.3 \mbox{
GeV}\mbox{ cm}^3$, and with a slope $\rho \propto r^{-2}$, the DAMA
result is compatible with the detection of a DM particle with a mass
around 50 GeV and a WIMP-nucleons scattering cross section of order
$10^{-41}-10^{-42} \mbox{ cm}^2.$

Other experiments, such as CDMS and EDELWEISS, have explored the
region of parameter space allowed by the DAMA modulation signal,
finding null results \cite{Edel null searches, CDMS null searches}.
The comparison between the DAMA annual modulation and the other
mentioned experiments is however model-dependent. Taking into
account astrophysical uncertainties, the DAMA allowed region is
sensibly increased, with masses extending up to $\sim 250$ GeV and
spin independent WIMP-proton cross section down to $10^{-43} \mbox{
cm}^2$ \cite{Belli LargeA, Bernabei LargeA, Belli2 LargeA}.
Nevertheless, null searches of recent experiments make the most
na\"ive interpretation of the DAMA signal problematic \cite{Copi,
CDMS2006, SI XENON}.

It should be stressed, however, that the DAMA signal should not be
dismissed without further investigation, especially in view of the
fact that theoretical scenarios (despite exotic) exist where an
interpretation in terms of DM appears not to be in conflict with
other existing experiments~\cite{Bottino:2007qg} (see also
references therein).

The upper bounds on the WIMP spin-independent coupling inferred by
several experiments are summarized in Fig.
\ref{fig:SpinIndependent}. The most stringent result (as of November
2007) was obtained  by the XENON collaboration. The limits on SD
cross section are far weaker and the best constraints, plotted in
Fig. \ref{fig:SpinDedependent}, come from CDMS \cite{Akerib SD},
NAIAD \cite{SD NAIAd}, Super-Kamiokande \cite{SD SuperKAM} and KIMS
\cite{SD KIMs}. A better sensitivity to spin dependent couplings is
expected for the COUPP experiment, a heavy liquid bubble chamber
under development in the NuMi gallery at Fermilab \cite{COUPP}.

In comparison, the theoretical predictions of neutralino elastic
scattering off nucleons, for different SUSY scenarios, show that
current direct searches have begun to explore a relevant portion of
the parameter space, while improved sensitivities are needed to
perform a complete scan \cite{Ellis NeutralinoDD, Bottino Neutr DD}.

Direct detection constraints exclude the left-handed sneutrino in
the MSSM as dominant Dark Matter component. However, the
right-handed sneutrino, in extensions of the MSSM, is a viable Dark
Matter candidate, compatible with direct searches \cite{LeeMatchev,
ArinaFornengo}.

\section{IS IT COMPATIBLE WITH GAMMA-RAY CONSTRAINTS?}
\label{sec:chapter eight}

Aside from direct and accelerator searches, one may search for
DM through the detection of its annihilation
products, such as photons, anti-matter and neutrinos.

In particular, since the energy scale of the annihilation photons is
set by the DM mass, and since some of the most studied DM candidates,
such as the supersymmetric neutralino and the LKP in UED models, are
expected to lie in mass in GeV-TeV region, exotic gamma-ray sources
are among the primary targets of indirect searches (see e.g. \cite{Bertone
Rev. gamma-ray} for a review about DM searches though gamma-ray
astrophysics).
Significant emissions at other wavelengths is however predicted in
most cases, due to
the interactions of the annihilation products with ambient photons
or magnetic fields, making multi-wavelength searches possible.
Emission at different energy scales has also been discussed in the
context of other DM candidates, e.g.
X-rays from the decay of sterile neutrinos (see Sec.\ref{sec:chapter nine}).

The gamma-ray flux from WIMP annihilations in a DM halo depends on the
particle physics parameters as well as on cosmological quantities,
such as the profile of DM halos. More precisely, the gamma-ray flux at earth
(if the WIMP is not its own antiparticle a factor 1/2 must
be added) is given by
$$ \phi(\psi,E_{\gamma})= \frac{\langle \sigma_{ann}v\rangle}{8\pi m_{\chi}^2} \frac{dN_{\gamma}}{dE}\times \int_{l.o.s.}ds \rho^2(r(s,\psi)), $$
where $m_{\chi}$ and $\langle \sigma_{ann} v \rangle$ are
respectively the mass and the cross section annihilation times
relative velocity of the DM particle. From Eq. \ref{eqn: Omega} in
Sec. \ref{sec:chapter one} it follows that, to match the correct DM
density, it is necessary for a cold thermal relic $\sigma_{ann} v
\sim 10^{-26} \mbox{ cm}^3 \mbox{s}^{-1},$ although this value is
just indicative because, for instance, coannihilations can
substantially modify the picture, plus the cross section in the non
relativistic limit may substantially differ from the one at
decoupling (e.g. in the case of p-wave annihilations).
$\frac{dN_{\gamma}}{dE}$ is the photon spectrum from DM
annihilations, that depends on the nature of DM candidate. Finally,
the last term in the equation is the integration along the line of
sight of the dark matter density squared. The quadratic dependence
on $\rho$,  suggests that ideal targets of indirect searches are
regions where the DM density is strongly enhanced such as the
Galactic center (e.g. \cite{Aharonian:2006wh, Zaharijas:2006qb,
Berezinsky, Bergstrom Ullio GC, Profumo:2005xd, Cesarini:2003nr,
Bouquet:1989sr,Bertone Merrit spikes2002}), halo substructures (e.g.
\cite{Silk:1992bh, Bergstrom:1998zs, Calcaneo-Roldan:2000yt,
Aloisio:2002yq, Koushiappas:2003bn, Diemand:2005vz}) and the core of
external galaxies (e.g. \cite{Baltz Briot, Evans Ferrer, Tyler,
Pieri:2003cq}). Prospects for detecting gamma-rays have been
discussed also for overdensities in DM halos called caustics (e.g.
\cite{hogan, Bergstrom caustics, Pieri Branchini caustics, Mohayaee
shandarin, Natarajan caustics}). Much steeper profiles, called {\it
spikes}, may form due to adiabatic growth of black holes, for
example around the Super Massive black hole at the Galactic center
(see e.g. \cite{Bertone:2005hw, Bertone Merrit Spike2005, Bertone
Merrit spikes2002} for a discussion about the prospect for detecting
DM annihilation gamma-rays in this scenario). Although in this case
the spike is likely disrupted by astrophysical processes
\cite{Bertone:2005hw, Merrit Verde, Ullio Zhao}, a moderate
enhancement, called crest, may form again due to gravitational
interactions with the observed stellar cusp \cite{Merrit Harsf
crest}. More promising targets may be mini-spikes around
intermediate massive black holes, since they are not affected by
dynamical processes that tend to lower the density enhancement
\cite{Bertone Zentner Silk, FornasaTaosoBertone}.

\begin{figure}[tb]
\includegraphics[width=8.5cm]{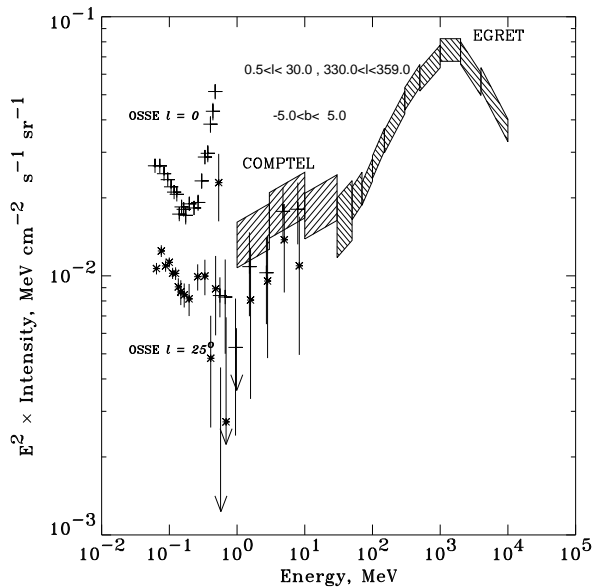}
\caption{Spectrum of the inner Galactic plane for $|b|<5^{\circ}$
measured by different experiments for energies ranging from sub MeV
up to tens of GeV. From Ref. \cite{Comptel}} \label{fig:SpectrumGP}
\end{figure}

Although conclusive evidence for Dark Matter annihilations has not
been obtained so far, gamma-ray experiments have nonetheless
provided a wide range of observations that can be used as upper
bounds of gamma-ray fluxes from DM annihilations, in order to
constrain existing DM scenarios. In particular observations in the
soft gamma-ray energy band, between roughly 50 keV and 1 MeV, have
been performed by the Osse experiments \cite{Osse Skibo Kinzer} and
more recently by INTEGRAL in the range 20-8000 keV (see
\cite{Weidenspointner, Weidenspointner06}). All-sky observations
have been performed by COMPTEL in the energy range 3 MeV - 30 MeV
\cite{Comptel}, and EGRET from 30 MeV to over 30 GeV \cite{Cillis
Hartman Egret}. In Fig.\ref{fig:SpectrumGP} we show the spectrum of
the inner Galactic plane as measured by these experiments.

Current Air Cherenkov Telescopes such as CANGAROO \cite{Cangoroo},
HESS \cite{hess}, MAGIC \cite{magic} and VERITAS \cite{veritas} are
collecting data at higher energies and the GLAST satellite
\cite{Glast}, which is scheduled for launch in 2008, will allow
much deeper observation in the energy range 20 MeV - 300 GeV.

As we have seen, there is no conclusive evidence of DM
annihilations, but many claims of discovery, or hints of detection,
have been been put forward in recent years. For example the
gamma-ray source been detected by EGRET in the direction of the
Galactic center, has been interpreted in terms of DM annihilations
(as discussed e.g. in \cite{Bouquet:1989sr, Berezinsky, Bergstrom
Ullio GC, Cesarini:2003nr, Fornengo Pieri Scopel}), although it was
subsequently suggested that the source may be slightly offset with
respect to the Galactic center \cite{Hooper Dingus}.

The HESS experiment has recently
discovered a very high energy source spatially coincident with Sgr
$\mbox{A}^*$, the compact radio source at the Galactic center, and
the spectrum has been subsequently confirmed by the MAGIC collaboration.
Even in this case, however, the bulk of the signal can hardly be interpreted in
terms of the annihilation of common DM candidates, since the shape of
the energy spectrum is close to a perfect power-law over two decades in energy,
a circumstance that rather points towards ordinary astrophysical
sources \cite{Aharonian:2006wh,Profumo:2005xd}.

Hints of a DM signal may hide in the extragalactic gamma-ray
background (EGB) which is is inferred from EGRET observations, after
subtraction of the galactic component(see e.g. \cite{Bergstrom Edsjo
Ullio, Taylor Silk, Ullio Bergstrom, Ando, Horiuchi, Ahn Bertone}).
The existence of a bump in the EGB spectrum at few GeV \cite{Hunter
et al, Strong Moska Reimer} has been tentatively interpreted in
terms of DM annihilations \cite{Elsasser Mannheim, De Boer 2004,
DeBoer Sander}. This cannot be however considered as evidence for
DM, since the freedom in the DM (cosmological and particle physics)
parameters allow enough freedom to explain almost any excess
observed in the GeV-TeV range (see the discussion in \cite{Bertone
freedom}). In order to obtain conclusive answers, a more robust
evidence could be provided by the power spectrum of the EGB
anisotropies as can be obtained e.g. by GLAST \cite{Ando Komatsu}.

Another observation awaiting for a (not necessarily exotic)
interpretation is the INTEGRAL detection of an intense 511 keV
emission line, due to positron annihilations, towards the galactic
center. Many astrophysical sources of positrons have been proposed,
for example interactions of cosmic-rays with the interstellar medium
\cite{Kozlowsky Loeb}, pulsars \cite{Sturrock}, gamma-ray bursts
\cite{Bertone Kusenko},  microquasars \cite{Guessoum Jean} or
radiactive nuclei expelled by stars such as supernovae, Wolf-Rayet
and red-giants \cite{Casse Cordier} (see \cite{Bertone Rev.
gamma-ray, Weidenspointner} and references therein). However,
conventional astrophysical scenarios, can hardly explain the size
and morphology of the emitting region, that coincides roughly with
the Galactic bulge and that exhibits a fainter disk component
\cite{Weidenspointner, Weidenspointner06}. Other more exotic
interpretations are again open, in particular the positron source
may be provided by DM annihilations. DM candidates with masses close
to the electroweak scale have been excluded because the concomitant
photon emission would violate the gamma-ray bounds. However, this
problem may be circumvented in models where the WIMP shares a
quantum number with a specie nearly degenerate in mass, with a
splitting in the MeV-range \cite{Pospelov Ritz, Finkbeiner Weiner}.

It has been shown that a DM candidate in the MeV range may
succesfully explain the 511 keV line, while remaining compatible
with other observational constraints \cite{Boehm Ensslin Silk}. A
list of alternative candidates include axinos \cite{Hooper Dingus},
sterile neutrinos \cite{Picciotto Pospelov}, cosmic strings
\cite{Ferrer Vachaspati}, moduli \cite{Kawasaki Yanagida moduli},
Q-balls \cite{Kasuya Takahashi} and scalars coupled to leptons with
gravitational strength \cite{Picciotto Pospelov}.

Upper limits on the WIMP mass, in order to be consistent with the
EGRET and COMPTEL bounds, can be derived by comparing the gamma-ray
emission from internal bremsstrahlung processes and in-flight
annihilations with existing gamma-ray data \cite{Beacom Bell
Bertone, Beacom Yusek, Sinuz Casse Schanne}, that set an upper limit
on the mass of the DM particle of about 3-7 MeV (but see also
\cite{Boehm Uwer}). This would be in conflict with the lower bound
on the MeV DM particle mass ($\sim 10$ MeV) inferred by the cooling
rate and neutrino emission of the SN 1987A \cite{Fayet SN}, unless
the coupling of these particles with neutrinos is suppressed.

Anyway, all the Dark Matter interpretations of the 511 keV line are
today disfavored by recent observations of the emission
\cite{IntegralNature}.

It is thus important to search for clear, smoking-gun signatures of
DM. The first, and maybe foremost, would be the detection of mono
energetic gamma-ray lines, produced e.g. by neutralino or LKP
annihilations via loop-diagrams with $\gamma \gamma $ or $\gamma
\mbox{Z}$ as final states (see e.g. \cite{Bergstrom Ullio GC, Rudaz
Stecker, Bergstrom Edsjo Ullio, Boudjema Semenov, Bergstrom Brigman
2gammaLKP}) or also in models with scalar dark matter \cite{Boehm
Orloff, GustafssonBergEdsjo}. A number of alternative strategies
have been proposed over the years, see e.g. Ref.
\cite{Bertone:2007ki} for a recent review.

\section{IS IT COMPATIBLE WITH OTHER ASTROPHYSICAL BOUNDS?}
\label{sec:chapter nine}

\subsection*{Neutrinos}

Neutrinos can be produced in DM annihilations either directly or
 via the  decay of other annihilation products, and may be
detected with high-energy neutrino telescopes, that measure
the Cherenkov light emitted by secondary muons propagating in
water or ice.

The Sun and the Earth have been proposed as targets for indirect DM
searches, since a large number of WIMPs could accumulate in their
interior, releasing a large number of neutrinos. The neutrino flux
depends on the capture rate of WIMPs in the Sun or in the Earth,
thus on the elastic cross section of these particles.

The spin-dependent cross section is far less
constrained than the spin-independent one (see
Sec.\ref{sec:chapter six})
Since in the Earth the abundance of nuclei with odd atomic numbers
is very small, the capture rate is dominated by the strongly
constrained spin independent coupling, contrary to what happens in
the Sun. The prospects for detecting neutrinos from the center of
the Earth are therefore not particularly promising, at least for
current and upcoming experiments \cite{Lundberg Edsjo}. The null
searches of AMANDA have been used to derive an upper limit on
neutrino flux from WIMP annihilations, for example in the framework
of neutralino Dark Matter \cite{Achterberg06}. In the framework of
MSSM, however, the most optimistic neutralino scenarios will be
probed by kilometer size neutrino telescopes such as IceCube
\cite{Bergstrom Edsjo Gondolo98}.

The prospects for detecting neutrinos from the annihilation of
Kaluza-Klein Dark Matter in the Sun are more promising, because
of its large axial coupling, and the annihilations of $B^1$ particles to
neutrino and tau leptons pairs, respectively forbidden and
subdominant in the case of neutralino, dominate the neutrino
spectrum, producing a large number of high energy neutrinos.
The event rate in kilometer scale detectors is
expected between 0.5 and 10 events per year \cite{Hooper Kribs03}.

The Galactic Center (GC), a well studied site for gamma-ray DM
searches, offers instead poor prospects for detecting Dark Matter
annihilations though neutrinos. An upper limit on neutrino flux from
GC, in the case of neutralino Dark Matter, has been obtained by
requiring that the associated gamma-ray emission would not exceed
the flux measured by EGRET \cite{Bertone Nezri04}. Unfortunately
this bound is below the sensitivity of present and upcoming experiments, such
as ANTARES, unless extreme scenarios are considered.

The prospects for detecting neutrinos from so-called {\it mini-spikes}
around Intermediate Mass Black
Holes (see the discussion in Sec. \ref{sec:chapter eight}) appear
more promising. The strong
enhancement of the DM density around these objects induces a
substantial boost of the DM annihilation rate, leading to
neutrino fluxes within the reach of ANTARES and IceCube \cite{Bertone IMBHnu06}.


A combination of data from different neutrino telescopes can
already be used to set an upper bound on the total DM annihilation
cross section in the non-relativistic limit, for WIMP masses
between 100 MeV and $10^{5}$ GeV, which is stronger than the unitarity
bound. \cite{Beacom Bell
Mack06}.

\subsection*{Antimatter}

Indirect searches of Dark Matter can be performed by looking at
an exotic contribution in the spectra of positrons and antiprotons
 in cosmic-ray fluxes.
These charged messengers, contrary to gamma-rays and neutrinos,
do not provide information on the location of their source
because of the interaction with the interstellar
magnetic fields.

Positrons in cosmic-rays mostly originate from
the decay of charged pions and kaons produced in cosmic-ray
interactions with interstellar gas.
Analytic treatments and numerical codes have been
developed to describe the propagation of cosmic rays, and to compute
the amount of secondary cosmic rays, including positrons, produced by
collisions of primary particles with the interstellar medium (see
e.g. \cite{Strong Moska Review} for a review about the cosmic-ray
propagation in the Galaxy). The measurement of the positron fraction,
i.e. the ratio of the positron flux over the sum of the positron and
electron fluxes, provides an interesting tool to search for exotic
positron sources in a region of a few Kpc from us.

In 1994, the HEAT experiment has observed an excess of the positron
fraction, with respect to standard propagation models, at
energies beyond 7 GeV \cite{Heat97}. This result has been
subsequently confirmed by further measurements obtained by HEAT
\cite{Coutu} and recently by re-analysis of the AMS-01 data
\cite{Olsem, Jacholkowska}. WIMP annihilations have been proposed to
explain this enhancement, in particular in the framework of
supersymmetry \cite{Batz Edsjo99, Baltz Edsjo Freese01, Kamionk
Turner91, Kane Wang02, Turner Wilczek90, Hooper Taylor Silk,
HooperSilkPositron, Brun:2007tn} and Kaluza-Klein Dark Matter
\cite{HooperSilkPositron, Hooper Kribs}. The spectral shape of the
positron excess can be well reproduced by LSP annihilation models
but a very high annihilation rate is necessary to match the correct
normalization, requiring therefore unnaturally large amounts of
local dark matter substructures \cite{Hooper Taylor Silk}. Instead,
a modest boost factor is needed for Kaluza-Klein Dark Matter,
due to their large annihilation rate to charged leptons,
that leads to a larger number, and harder spectrum, of positrons
\cite{HooperSilkPositron, Hooper Kribs}.

\begin{figure*}[tb]
\includegraphics[width=6.5cm]{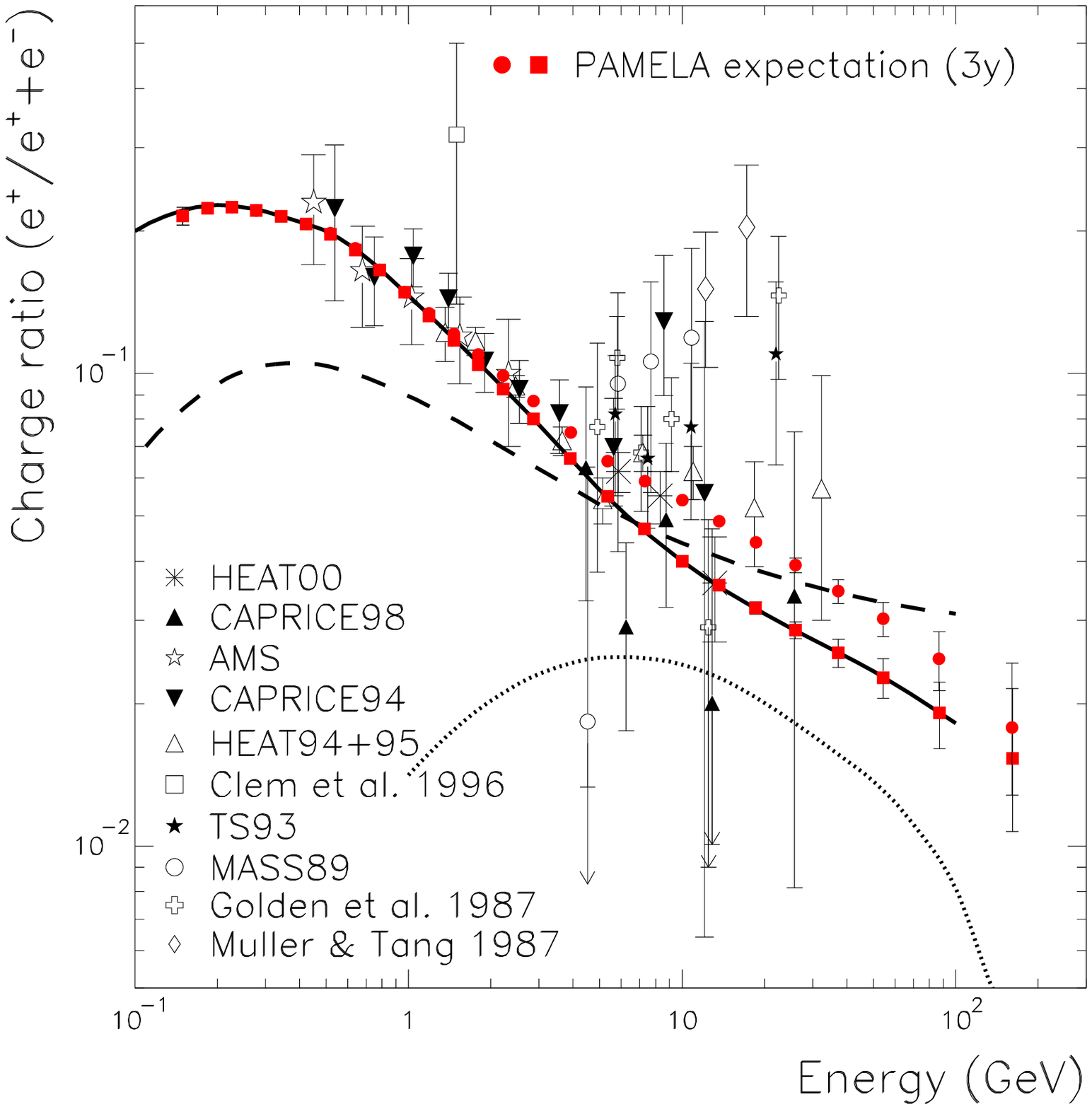}
\includegraphics[width=6.5cm]{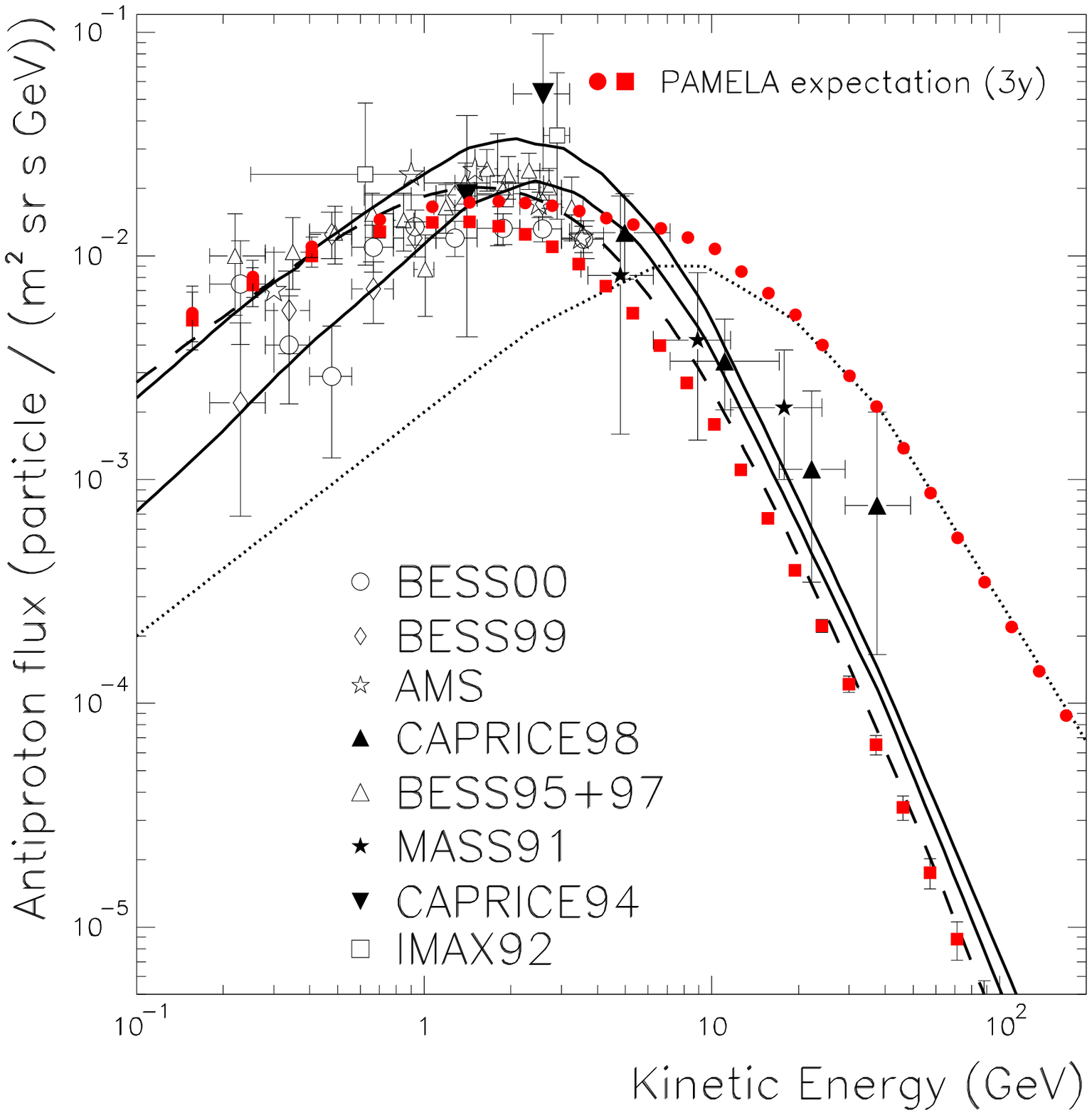}
\caption{Left: Positron fraction as a function of energy. Shown are
the theoretical calculations for pure secondary production (solid
and dashed lines) and for pure primary production from neutralino
($m_{\chi}$=336 GeV) annihilations (dotted line). Red circles
(squares) show a projection of the Pamela measurements after three
years of data taking with (without) a neutralino contribution.
Right: Energy spectrum of anti-protons. Shown are the theoretical
prediction for secondary (solid and dashed lines) and primary
production from neutralino ($m_{\chi}$=964 GeV) annihilations
(dotted line). From Ref.\cite{Picozza}.} \label{fig:antimatter}
\end{figure*}

The measurements of the positron fraction performed by several
experiments, including AMS-01, CAPRICE and HEAT are in agreement
with each other but the large uncertainties, as shown in
Fig.\ref{fig:antimatter}, do not allow to draw definitive
conclusions on the nature of the GeV excess. However, the situation
may be clarified soon thanks the larger positron statistics that
will be obtained by the PAMELA satellite \cite{Picozza}, which was
launched in June 2006, and the AMS-02 experiment \cite{AMS-02},
which should be launched in the near future.

Dark Matter annihilations in the galactic halo may produce
antiprotons and the possible imprint
on cosmic rays measurements, both at low and high energies, has
been extensively studied \cite{Donato Fornengo2003, BottinoDonato05,
BarrauSalatiDonato, Stecker85, MoreselliLionetto, HisanoMatsumoto,
Bringman Salati06}. However the data collected by several
experiments, in particular BESS, CAPRICE and BESS-Polar, agree with
the calculations of the positron production by cosmic rays, showing
no evidence for primary antiprotons \cite{Brigman Edsjo Ullio99}
(Fig.\ref{fig:antimatter} shows data from recent experiments and
also theoretical calculations for pure secondary/primary antiproton
production). Furthermore, these results cannot be easily translated
into constraints on DM candidates, due to the large uncertainties on the
antiproton flux induced by the propagation parameters \cite{Donato
Fornengo2003, BarrauSalatiDonato}. Much more data will soon be available
thanks to Bess-Polar, PAMELA and
AMS-02, in particular at high energies, where the contribution from
DM annihilations might be dominant for heavy enough WIMPs \cite{Bringman Salati06}.

Despite the uncertainties in the propagation models, the study of
anti-matter fluxes can sometimes provide a useful diagnostic tool
for specific DM scenarios. For instance,
Bergstr$\ddot{\mbox{o}}$m et al. have investigated the DM
annihilation model of Ref.\cite{de Boer Sander Glady05},
proposed to explain the EGRET excess of the diffuse galactic gamma
ray background, by computing the associated primary antiproton flux
from WIMP annihilations. They were then able to rule out the scenario
since the anti-proton flux was found to grossly exceed the measured
anti-proton flux \cite{Bergstrom Edsjo
Gustafsson06}. The model of Ref.\cite{de Boer Sander Glady05} might
still be made compatible with observations allowing for an
anisotropic diffusion of cosmic rays \cite{deBoer Gebauer06}.

\subsection*{Multi-wavelength approach and X-Rays emission}

More in general, a multi-messenger, multi-wavelength analysis
provides more robust results than the simpler {\it fit-the-bump}
approach. For example, when interpreting the origin of a gamma-ray
source, one may study the associated synchrotron, bremsstrahlung and
Inverse Compton emission, produced by electrons and positrons
inevitably produced along with gamma-rays in DM annihilations. These
signals can in principle extend over a wide range of wavelengths,
all the way from radio to gamma-rays. The limited field of view of
radio and X-ray experiments makes it easier to perform
multi-wavelength studies of a restricted number of candidate
sources, such as the Galactic center \cite{Bertone Merrit
spikes2002, AlosisioBlasi, PieriBergstrom} galaxy clusters
\cite{ColafrComa, ColafrBernCluster}, and dwarf galaxies
 \cite{BaltzWai, ColafrancescoDraco}. Radio and X-ray
observations are powerful techniques to search for WIMP
annihilations and they can provide constraints even more restrictive
than those inferred from gamma-rays \cite{ColafrancescoDraco,
PieriBergstrom}.

X-ray observations provide useful constraints also on DM candidates
other than WIMPs. For example, sterile neutrinos (see Sec.
\ref{sec:chapter two}) can decay into active neutrinos
$\nu_{\alpha}$ and photons with energies in the X band: $\nu_s
\rightarrow \nu_{\alpha} + \gamma,$ $\mbox{E}_{\gamma} = m_s/2.$
Therefore, X-rays observations constrain the sterile neutrino mass
$m_s$ and their mixing angle with active neutrinos. Assuming then a
production mechanism (see e.g. \cite{Abazajan Production}), these
limits can be turned into an upper bound on the particle mass. The
observation of the cosmic X-ray background requires $m_s < 8.9$ keV
(95 \% C.L.) \cite{Boyarsky05}, but more stringent constraints are
obtained from individual objects, such as galaxies or clusters of
galaxies. For example the XMM-Newton observations of Virgo A impose
$m < 10.6$ keV (95 \% C.L.) \cite{Boyarsky06} and an analysis of the
Virgo and Coma cluster data further restrict the bound to $m < 6.3$
keV (95 \% C.L.) \cite{Boyarsky06, Abazajan Koushiappas06}. A
significant improvement has been obtained from X-ray observations of
the Andromeda Galaxy: $m < 3.5$ keV (95 \% C.L.) \cite{Watson
Beacom06}. These results, combined with the lower limit on the
sterile neutrino mass inferred by measurements of small scale
clustering (see Sec. \ref{sec:chapter two}), rule out sterile
neutrinos in this scenario as the dominant Dark Matter component,
constraining their fraction on the total Dark Matter amount to be
$f_s \lesssim 0.7$ at the 2 $\sigma $ level \cite{Palazzo}.

However, sterile neutrinos remain viable for alternative production
mechanisms, such as Higgs decays in models with an extended Higgs
sector \cite{Kusenko and Petraki}, or a resonant production in
presence of a very large lepton asymmetry in the Universe, $\mbox{L}
>> 10^{-10}$ \cite{Abazajian Fuller,Resonant Sterile}.

\section{CAN IT BE PROBED EXPERIMENTALLY?}

The last requirement for a particle to be a good DM candidate,
is that such particle can be probed experimentally, in the sense that
it can be directly detected or that convincing evidence for it,
or for the theoretical scenario it arises from, can be obtained
with present or future experiments.
The nature of this requirement is different from that of the nine other
conditions discussed above, where we have essentially required
that DM scenarios are not in conflict with existing experiments and
observations. Here we add the requirement of "discoverability",
that reflects our prejudice on what can be
considered a {\it good} theory in science.


\subsection{Probing SuperWimps}

DM particles may interact far less than weakly, and they could
evade all conventional dark matter searches. For example, the
supersymmetric gravitino, which only couples gravitationally,
 has been proposed as a well motivated Dark-Matter candidate.
The LKP graviton in UED, axions and axinos are other examples of
super-weakly interacting massive particles (or super-WIMPs), i.e.
Dark Matter candidates that can be extremely difficult or impossible
to observe in direct and indirect Dark Matter searches because of
their very suppressed interactions \cite{Feng Rajaraman03,
ChoiRoszkowski05, Cembranos Feng07}.

However, the next to lightest supersymmetric
particle (NLSP) could be long-lived, for example the stau NLSP
lifetime is of order $10^6$ sec, for gravitino masses of $10$ GeV
\cite{HamaguchiKuno04}. If the NLSP is a neutralino, this scenario
may have an interesting collider
signature, similar to the case of neutralino LSP,
because the decays of the NLSP neutralino may
occur outside the detector. In this case, the sparticle spectrum
may allow the discrimination between gravitino and neutralino LSP
through the analysis of selected decay channels \cite{DeRoeck Ellis07},
or spins \cite{BuchmullerHamaguchiPLB04}, even if this programme may
be challenging for the LHC.

A stau NSLP scenario, as possibly realized in supergravity models
\cite{DeRoeck Ellis07}, offers a more promising opportunity to
uncover gravitino Dark Matter models at colliders. The
charged NLSP particle particle would have distinctive time-of-flight
and energy-loss signatures that might enable to reconstruct its mass
with high accuracy, at a level of per cent or even smaller
\cite{DeRoeck Ellis07, EllisRaklev06, HamaguchiNojiri07}. A stau
would be produced at the end of every supersymmetric cascade and
being strongly ionizing, if it is sufficiently slow moving, it may
be stopped inside the detector or in a surrounding water tank or
calorimeter detector. In particular, it has been suggested that up to
$\mathcal{O}(10^3)$ and $ \mathcal{O}(10^4)$ charged NLSP can be
trapped per year at LHC and ILC respectively, by placing a 10 Kton
trap around the detector \cite{HamaguchiKuno04, FengSmith05,
HamaguchiNojiri07}.

Collecting a large number of stau, it would be possible to measure
the stau lifetime and kinematically determine the gravitino mass,
from the dominant decay $\tilde{\tau}\rightarrow \tau + \tilde{g}.$
The measurements of gravitino and stau masses would allow to compute
the stau lifetime predicted by the supergravity model and if it
matches  the experimental value, one would obtained a strong
evidence for supergravity and for gravitino LSP
\cite{BuchmullerHamaguchiPLB04, HamaguchiKuno04, FengSmith05}.

Detailed simulations have been performed to study the
gravitino Dark Matter scenario at LHC and ILC (e.g.
\cite{DeRoeck Ellis07, EllisRaklev06, Martyn06}). In particular, at
ILC, with an integrated luminosity of 200 fb$^{-1}$ at $\sqrt{s} =
420$ GeV, thousands of stau will be stopped within the hadron
calorimeter, allowing a reconstruction of the gravitino mass with an
accuracy of few GeV and a determination of the Planck mass, for a
test of supergravity predictions, at a level of 10 \%
\cite{Martyn06}.

As discussed in Sec.\ref{sec:chapter four}, for NLSP and gravitino
masses in the GeV range, BBN bounds severely constrain the case of
neutralino and stau NLSP, while a sneutrino NLSP is perfectly
viable. In the latter case, the NLSP decay is invisible, but the
predicted small sneutrino-stau mass splitting may produce
interesting collider signatures, with soft jets or leptons in the
final states \cite{CoviKraml07}. Finally, models of gravitino DM
with broken R-parity, may also be searched for in accelerators
\cite{Covi, HirschPorod05, LolaOsland},  but also through indirect
detection \cite{Bertone:2007aw,Ibarra:2007wg}.

\subsection*{Axinos}

Axinos appear in supersymmetric models implementing the Peccei-Quinn
mechanism for solving the strong CP problem, and  correspond to the
fermionic superpartner of the axion. Their mass ranges between the
eV and the GeV scale, and they can be
efficiently produced through thermal and non-thermal processes in the
early Universe under the form of cold, warm or even hot Dark matter (see e.g. Refs.
\cite{CoviRoszkowski02Axino, ChoiRoszkowski05, Steffen05Axino,
CoviRoskowski04Axino} and references therein).

In particular, axinos cold Dark Matter is achieved for masses $m\geq
100$ keV and for low reheating temperatures $T_R \leq 10^{6}$ GeV,
in contrast with gravitino CDM that can allow for higher values of
$T_R$, such as $T_R \sim 10^{10}$ GeV for $m_{\tilde{g}} \sim 1$
TeV.

Axino couplings are suppressed by the inverse of the Peccei-Quinn
breaking scale, $f_{a} \geq 10^{9}$ GeV, and therefore these
particles are extremely weakly interacting. As a consequence,
similarly to the case of gravitino LSP, the lifetime of the NLSP can
be long, and the strong bounds from Big Bang Nucleosynthesis avoided
\cite{CoviRoskowski04Axino}.

The direct production of axinos at colliders is strongly suppressed
but they can be profusely produced by the decays of the NLSP
particles. As in the case of gravitino, a large number of sleptons
NLSP could be collected, in order to measure the NLSP lifetime and
to reconstruct the axino mass and the Peccei-Quinn scale $f_a$ (see
\cite{Steffen05Axino} and references therein).

However, the problem may arise of discriminating between axino and
gravitino LSP models. For instance, a stau NLSP with a lifetime
within the range 0.01 s - 10 h is predicted in both scenarios, while
shorter or longer lifetime are possible only with a gravitino LSP.
To solve this ambiguity, one may consider the three body decay
$\tilde{\tau} \rightarrow \tau + \gamma + \tilde{g}/\tilde{a}.$ For
at least $\mathcal{O}(10^4)$ observed stau decays, a clear
distinction between the two models can be achieved through the
angular distribution of the decay products and/or measuring its
branching ratio \cite{BrandeburgCovi05Axino, HamaguchiNojiri07}.

\subsection*{Axion}

Axions have been proposed as a viable CDM candidate and the
suppression of their interactions by the Peccei-Quinn scale makes
them very weakly interacting (see e.g. \cite{Sikivie,Raffelt Axions}
and references therein for more informations). Their relic abundance
matches the Dark Matter cosmological density for masses around 10
$\mu$eV but significant deviations from this value can occur because
of the large uncertainties in the production mechanisms.

One of the most prominent phenomenological properties is the two
photon interaction that allows axion-photon conversions in presence
of an electromagnetic field:
$$ \mathcal{L}_{a\gamma} = g_{a\gamma} \bold{E}\cdot\ \bold{B}\mbox{
}a.$$ Here, $\bold{E}$ and $\bold{B}$ are respectively the electric
and magnetic fields, $a$ is the axion field and $g_{a\gamma}$ is the
coupling constant.

This coupling constant is linearly related to the mass of the axion
and connected to measured properties of the pions and to details of
the underlying particle physics model. However, in some cases this
relation can be relaxed postulating the existence of axion-like
particles with unconnected masses and couplings.
The Primakoff process that converts axions into photons is at the
basis of most axion searches (see e.g. \cite{BattestiAxion07,
Raffelt Axions} for a discussion about axion searches).

For example, galactic Dark Matter axions could be resonantly
converted into microwave photons in the magnetic field permeating a
cavity. The signal would carry information on the mass as well as
the axion distribution in the galactic halo. The ADMX
experiment\cite{AsztalosADMX} has already started to explore the
region of parameter space favored for Dark Matter axions, while
a larger portion will be probed by the upgraded
version of the same project and upcoming microwave cavity
experiments.

Complementary searches are dedicated to axions produced by photon
conversion in the electromagnetic field of the Sun, probing regions
of the parameter space where axions are unlikely the dominant
component of Dark Matter.

Solar axions can be searched for with axion helioscopes, through the
reconversion to X-rays in external magnetic fields. The strongest
bound is obtained by the null searches of the CAST experiment
$g_{a\gamma} < 8.8\cdot 10^{-11} \mbox{ GeV}^{-1}$ for
$m_a\lesssim0.02 \mbox{ eV}$ \cite{AndriamonjeCAST}.

In addition, one may look for the axion
Primakoff conversion into photons in the intense Coulomb field of
nuclei in a crystal lattice. However the inferred limits are less
restrictive with respect to the previous strategy.

Axions could also affect the polarization of a laser beam
propagating through a magnetic field. If the light is linearly
polarized with a non vanishing angle with the magnetic field
direction, the polarization plane rotates because the polarization
component parallel to the magnetic field is depleted by the
photon-axion conversion processes whereas the perpendicular
component does not.
In addition to the rotation of the polarization plane (dichroism),
an ellipticity is developed (birefringence) because of the different
refractive indexes of the parallel and transverse polarization
components.
A positive signal of dichroism and birefringence was initially
claimed by the PVLAS collaboration \cite{Zavattini} and some models
have been proposed to reconcile an axion-like interpretation
with existing astrophysical bounds \cite{Masso Redondo,
massoRedondo, MohapatraNasri07, JackelMasso07}. However, recent
observations, after an upgrade of the apparatus, appear to
suggest that the signal was likely due to instrumental
artefacts \cite{PVLas2007}.

Axions can also be searched for with photon regeneration
experiments, such as ALSP \cite{EhretALSP}. A laser beam propagates
through a magnetic field where photons can be converted into axions.
These particles, contrary to photons, can easily pass a opaque
bareer wall and they can be subsequently reconverted in photons by
the use of a second magnetic field. Finally, gamma-ray experiments
could be sensitive to axion-like particles because the photon-axion
conversions in the galactic magnetic field or in the photon
production sites could induce detectable signatures in the spectra
and fluxes of high-energy gamma ray sources
\cite{MirizziRaffeltSerpico, HooperSerpico07, DeAngelis Roncadelli,
DeAngelis Roncadelli2}.

\subsection{UED or SUSY?}

\begin{figure}[tb]
\includegraphics[width=8.5cm]{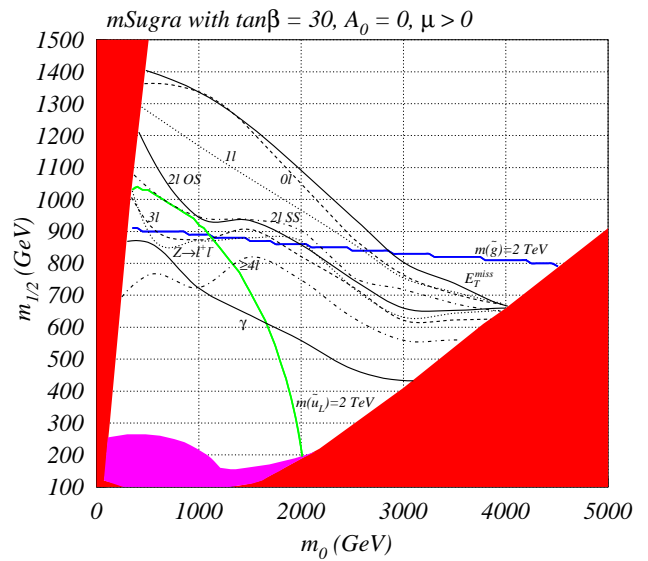}
\caption{An example of the reach of LHC to TeV-SUSY for different
channels in the plane $m_0$ vs $m_{1/2}$ in mSUGRA model. The
channels taken in account are: zero leptons (0l), one lepton (1l),
leptons with opposite charge (OS), leptons with same charge (SS),
three leptons (3l), four or more leptons ($\geq$4l), any numer of
leptons plus one photon ($\gamma$), at least 2 opposite sign leptons
with the invariant mass within an optimized interval around the $Z$
mass ($Z \rightarrow l^+ l^-$) and the inclusive missing transverse
energy channel. The solid lines are the 2 TeV mass contours for
squark and gluinos. The red region is excluded by theoretical
arguments and the magenta region is excluded experimentally. An
integrated luminosity of 100 fb$^{-1}$ is assumed. From Ref.
\cite{Baer Balazs03}} \label{fig:LHCReach}
\end{figure}

We conclude this section with a comment on the discrimination
between different DM scenarios. In fact, even in the case of the
most well studied DM candidates, i.e. the supersymmetric neutralino
and the LKP in UED, the experimental signatures may not easily allow
an unambiguous identification. As we have seen, neutralinos could be
pair-produced at LHC and escape the detector leading to an imbalance
of measured momentum. The discovery reach depends on the rate of
such missing energy events, that is strongly related to the squarks
and gluino masses. The discovery potential of LHC and the ability to
determine the SUSY parameters and masses for given supersymmetric
models have been extensively studied  \cite{Tovey02, Baer Balazs03,
ATLAS, CMS, Gjelsten04, Gjelsten05, Lester, DattaDas07, Baer07} and
for squarks and gluino lighter of 1 TeV the necessary integrated
luminosity will be available at LHC already in the first year of
operation \cite{Tovey02}. In Fig. \ref{fig:LHCReach} we show the
reach of LHC to TeV scale SUSY, for different channels.

An important role, in the discovery and understanding of SUSY, may
be played by the planned positron-electron International Linear
Collider (ILC) that should allow a more precise reconstruction of
the supersymmetric parameters.

The interplay of LHC and ILC
might be crucial for Dark Matter studies, because it would allow to
measure the particle physics cross sections and sparticle masses
with enough accuracy to infer the neutralino relic density and to
test whether the LSP really constitutes the Dark Matter \cite{Baltz
Battaglia 06} (see also \cite{LHC/ILC} for a broader discussion on the
complementarity of LHC and ILC).

The prospects for discovery of Universal Extra Dimension at LHC are
also promising. The most abundantly produced states are those
strongly interacting, i.e. the first level quarks and gluons, with
very large production cross sections for masses in the range of few
hundreds GeV \cite{MacesanuMvMullen02}. The first excitation of the
hypercharge gauge boson, $B^1$, can be the LKP and, thanks to KK
parity conservation, a good Dark Matter candidate. Similarly to
R-parity conserving SUSY, the first level KK states have to be pair
produced, and they subsequently decay into SM particles and into
$B^1$ LKP, with the latter escaping from the detector and leading to
a missing energy signature. At LHC, the signature with the largest
rate is $E^{miss}_T + (N\geq2) \mbox{ jets}$, but a more promising
channel for UED discovery is that of multilepton final states, with
the signature $4 l + E^{miss}_T.$ The  LHC should then probe an
inverse compactification radius of $R^{-1} \simeq 1.5 \mbox{ TeV}$
\cite{Cheng Marchev02}.

However, if a signal of new physics will be detected at LHC,
the problem will arise of discriminating between UED and SUSY
\cite{Cheng Marchev02}. In addition, also restricting to SUSY
models, LHC will leave degeneracies in the parameter space as it has
been shown in Ref. \cite{Arkani Kane Thaler, Binetruy04}.
Some specific features may simplify the task \cite{BattagliaDatta05, Datta Kong
Matchev05, SmillieWebber, AlvesEboli}, also for a discrimnation
between SUSY and Little Higgs model \cite{DattaDey07}.

For example, the spins of KK states are the same as their SM partners
while in SUSY they differ by 1/2. The spin determination at LHC will
be an extremely difficult task, but a charge asymmetry in the
lepton-jet invariant mass distributions from particular cascade
decays could be used to discriminate UED and SUSY. In particular,
quasi-degenerate mass spectra, such as those expected in UED, tend
to wash out the spin-correlations and therefore the prospect to
exclude a UED pattern given a SUSY spectrum are much better than
vice-versa \cite{Datta Kong Matchev05, SmillieWebber, AlvesEboli}.

Another difference between the two models is the structure of the
Higgs sector: in the Minimal UED model the analogues
of the heavy Higgs bosons in MSSM, $H^0,A^0,H^{\pm}$ are absent. Even if the
first level of Higgs boson has the same quantum numbers, it
appears more similar to the higgsino, since it carries KK parity.
However, this is not a robust criterium of discrimination at LHC
because there are regions of the SUSY parameter space where only the
SM higgs bosons can be detected, and therefore SUSY and UED could
be confused \cite{Cheng Marchev02}.

A smoking gun signature of UED models is instead provided by the
detection of second level particles. The $\gamma^2$ and $Z^2$ offer
the best prospect for discovery and their resonances can be
separately detected for $R^{-1} \leq 1 \mbox{ TeV}$ \cite{Datta Kong
Matchev05}. However, this is not a probe of UED because the
resonance could be interpreted as an extra $Z$ boson. A
quasi-degenerate $B^1$-$Z^1$ double resonance is instead a more
robust feature of UED, being an accidental mass degeneracy of
extra-$Z$ bosons unmotivated. However, this double peak structure
would be very difficult to observe at LHC (see Ref.\cite{Datta Kong
Matchev05} for more details).

In conclusion, it is likely that LHC alone will leave the door open
to several models. Anyway, ILC, with $\sqrt{s}=3 \mbox{ TeV},$ will
provide a more adequate tool to effectively distinguish UED and
SUSY. In particular the angular distribution of the events and the
threshold shape in the KK muons/smuons pairs production are the most
convincing evidences for UED/SUSY discrimination
\cite{BattagliaDatta05}.


\section{Conclusions}

In conclusion, we have presented a set of requirements that a
particle has to fulfil in order to be considered a viable DM
candidate. The requirements are presented in the form of a ten-point
test, and we have discussed each of them in a dedicated section that
describes the nature of the requirement and guides the reader
through the relevant literature.

The test performance of a small subset of DM candidates proposed
over the years is shown in Tab.\ref{tab:tabella}. The \checkmark
symbol is used when the candidates satisfy the corresponding
requirement, and it is accompanied by a ! symbol, in the case that
present and upcoming experiment will soon probe a significant
portion of the candidate's parameter space. If the requirement can
be satisfied only in less natural, or non-standard scenarios, or in
the case of tension with observational data, the symbol $\sim$  is
used instead. Candidates with a $\sim$ symbol in the last column,
where the final result is shown, should still be considered viable.
If one of the requirements is not satisfied, then the symbol
$\times$ is used, and since these requirements are {\it necessary}
conditions, the presence of a single $\times$ is sufficient to rule
out the particle as a viable DM candidate.

\begin{table*}
\begin{center}
\begin{tabular}{|p{5 cm}||*{11}{c|}|}
\hline
 & {\bf I.} & {\bf II.} & {\bf III.} & {\bf IV.} & {\bf V.} & {\bf VI.} & {\bf VII.} & {\bf VIII.} & {\bf IX.} & {\bf X.}& {\bf Result}\\
 {\bf {\it DM candidate}}  &   $\Omega h^2$   & Cold & Neutral & BBN & Stars & Self & Direct & $\gamma$-rays & Astro & Probed &  \\
                  \hline
\hline

SM Neutrinos & $\times$ & $\times$ & \checkmark & \checkmark & \checkmark & \checkmark & \checkmark & -- & -- & \checkmark &$\times$ \\
\hline
{\bf Sterile Neutrinos} & $\sim$ & $\sim$ & \checkmark & \checkmark & \checkmark & \checkmark & \checkmark & \checkmark & \checkmark ! & \checkmark & $\sim$ \\
\hline
{\bf Neutralino} & \checkmark & \checkmark & \checkmark & \checkmark & \checkmark & \checkmark & \checkmark ! & \checkmark ! & \checkmark ! & \checkmark & \checkmark \\
\hline
{\bf Gravitino} & \checkmark & \checkmark & \checkmark & $\sim$ & \checkmark & \checkmark & \checkmark & \checkmark & \checkmark & \checkmark & $\sim$ \\
\hline
{\bf Gravitino (broken R-parity)} & \checkmark & \checkmark & \checkmark & \checkmark & \checkmark & \checkmark & \checkmark & \checkmark & \checkmark & \checkmark & \checkmark \\
\hline
Sneutrino $\tilde{\nu}_{L}$ & $\sim$ & \checkmark & \checkmark & \checkmark & \checkmark & \checkmark & $\times$ &\checkmark  ! & \checkmark ! & \checkmark & $\times$ \\
\hline
{\bf Sneutrino} $\tilde{\nu}_{R}$ & \checkmark & \checkmark & \checkmark& \checkmark & \checkmark & \checkmark & \checkmark ! & \checkmark ! &\checkmark  ! & \checkmark & \checkmark \\
\hline
{\bf Axino} & \checkmark & \checkmark & \checkmark & \checkmark & \checkmark & \checkmark & \checkmark & \checkmark & \checkmark & \checkmark & \checkmark \\
\hline
{\bf SUSY Q-balls} & \checkmark & \checkmark & \checkmark & \checkmark & $\sim$& -- & \checkmark ! & \checkmark & \checkmark & \checkmark  & $\sim$\\
\hline
{\bf $B^1$ UED} & \checkmark & \checkmark & \checkmark & \checkmark & \checkmark & \checkmark & \checkmark ! & \checkmark !& \checkmark ! & \checkmark & \checkmark \\
\hline
First level graviton UED & \checkmark & \checkmark & \checkmark & \checkmark & \checkmark & \checkmark & \checkmark & $\times$ & $\times$ & \checkmark & $\times$$^a$
\\
\hline
{\bf Axion} & \checkmark & \checkmark & \checkmark & \checkmark & \checkmark & \checkmark & \checkmark ! & \checkmark & \checkmark & \checkmark & \checkmark \\
\hline
{\bf Heavy photon (Little Higgs)} & \checkmark & \checkmark & \checkmark & \checkmark & \checkmark & \checkmark & \checkmark & \checkmark !& \checkmark ! & \checkmark & \checkmark \\
\hline
{\bf Inert Higgs model} & \checkmark & \checkmark & \checkmark & \checkmark & \checkmark & \checkmark & \checkmark & \checkmark ! & -- & \checkmark & \checkmark \\
\hline
Champs & \checkmark & \checkmark & $\times$ & \checkmark & $\times$ & -- & -- & -- & -- & \checkmark & $\times$ \\
\hline
{\bf Wimpzillas} & \checkmark & \checkmark & \checkmark & \checkmark & \checkmark & \checkmark & \checkmark & \checkmark & \checkmark & $\sim$& $\sim$ \\
\hline
\end{tabular}
\end{center}
\caption{Test performance of selected DM candidates. The \checkmark
symbol is used when the candidates satisfy the corresponding
requirement, and it is accompanied by a ! symbol, in the case that
present and upcoming experiment will soon probe a significant
portion of the candidate's parameter space. If the requirement can
be satisfied only in less natural, or non-standard scenarios, or in
the case of tension with observational data, the symbol $\sim$  is
used instead. Candidates with a $\sim$ symbol in the last column,
where the final result is shown, should still be considered viable.
If one of the requirements is not satisfied, then the symbol
$\times$ is used, and since these requirements are {\it necessary}
conditions, the presence of a single $\times$ is sufficent to rule
out the particle as a viable DM candidate. Footnotes: $^a$ It is
possible to reconcile a graviton LKP scenario with CMB and diffuse
photon background measurements, if the minimal UED model is extended
with right-handed neutrinos, Ref.\cite{MatsumotoSato}.}
\label{tab:tabella}
\end{table*}


\section*{Acknowledgements}
We acknowledge partial support from the European Research Training
Networks HTRN-CT-2006-035863 (UniverseNet) and MRNT-CT-2004-503369
(Quest For Unification), the Special Research Project PRIN
2006023491 (Fundamental Constituents of the Universe) of the Italian
Ministry of University and Research (MiUR) and the INFN Special
Project on Astroparticle Physics FA51. One of us (M.T.) would like
to thank the International Doctorate on Astroparticle Physics
(IDAPP) of the MiUR for partial support during his stay at the IAP
in Paris and the IAP for hospitality.

\end{document}